\newif\ifconfver
\confvertrue       
\ifconfver
\documentclass[10pt,twocolumn,twoside]{IEEEtran}
\else
\documentclass[11pt,draftcls,onecolumn]{IEEEtran}
\fi

\usepackage{makecell}
\usepackage{bbding}
\usepackage{color,graphicx}
\usepackage{float}
\usepackage{multirow,amsmath,epsfig,amsfonts,amsbsy,amssymb,psfig,graphics,psfrag,theorem,calc,url,bm,cite}
\usepackage{stfloats,hyperref}
\usepackage{algorithm,algorithmic}
\usepackage{diagbox} 
\usepackage{hhline}
\usepackage{subfigure}
\usepackage{comment}
\usepackage{booktabs}
\usepackage{tikz}
\usepackage{quantikz}
\usepackage{pdfrender}

\usepackage{mathtools}

\makeatletter
\def\multilimits@{\bgroup
	\Let@
	\restore@math@cr
	\default@tag
	\baselineskip\fontdimen10 \scriptfont\tw@
	\advance\baselineskip\fontdimen12 \scriptfont\tw@
	\lineskip\thr@@\fontdimen8 \scriptfont\thr@@
	\lineskiplimit\lineskip
	\vbox\bgroup\ialign\bgroup\hfil$\m@th\scriptstyle{##}$\hfil\crcr}
\def\Sb{_\multilimits@}
\def\endSb{\crcr\egroup\egroup\egroup}
\makeatother

	{\end{list}}

\newlength{\twidth}
\ifconfver
\else
\fi

\definecolor{orange}{RGB}{255,107,0}

\newtheorem{Theorem}{Theorem}

\theorembodyfont{\rmfamily}

{\begin{list}{}{
			\settowidth{\labelwidth}{\mbox{\textnormal{#1}}}%
			\setlength{\leftmargin}{\labelwidth+\labelsep}}}%
	{\end{list}}

\newcommand\bA{\ensuremath{{\bm A}}}

\newcommand\bI{\ensuremath{{\bm I}}}

\newcommand\bM{\ensuremath{{\bm M}}}

\newcommand\bS{\ensuremath{{\bm S}}}

\newcommand\bX{\ensuremath{{\bm X}}}
\newcommand\bY{\ensuremath{{\bm Y}}}

\newcommand\bv{\ensuremath{{\bm v}}}

\definecolor{orange}{RGB}{255,107,0}

\ifconfver
\author{Chia-Hsiang Lin,~\IEEEmembership{Senior Member,~IEEE}, and Si-Sheng Young,~\IEEEmembership{Student Member,~IEEE}}

\else

\fi

\title{{HyperKING: Quantum-Classical Generative  Adversarial Networks for Hyperspectral\\ Image Restoration}

 \thanks{This study was supported by the Emerging Young Scholar Program (namely, the 2030 Cross-Generation Young Scholars Program) of National Science and Technology Council (NSTC), Taiwan, under Grant NSTC 113-2628-E-006-003.
  We thank the National Center for Theoretical Sciences (NCTS) and the National Center for High-performance Computing (NCHC) for providing the computing resources.}
    
   \thanks{\textit{(Corresponding author: Chia-Hsiang Lin)}}
\thanks{C.-H. Lin is with the Department of Electrical Engineering, and with the Miin Wu School of Computing, National Cheng Kung University, Tainan 70101, Taiwan (R.O.C.) 
(e-mail: chiahsiang.steven.lin@gmail.com).}
 \thanks{S.-S. Young is with the Institute of Computer and Communication Engineering, Department of Electrical Engineering, National Cheng Kung University, Tainan, Taiwan (R.O.C.)
		(e-mail:  q38121509@gs.ncku.edu.tw).}

}

\begin{document}
	
    \bibliographystyle{IEEEtran}
    \maketitle
    \ifconfver \else \vspace{-0.5cm}\fi

\begin{abstract}
Quantum machine intelligence starts showing its impact on satellite remote sensing (SRS).
Also, recent literature exhibits that quantum generative intelligences encompass superior potential than their classical counterpart, motivating us to develop quantum generative adversarial networks (GANs) for SRS.
However, existing quantum GANs are restricted by the limited quantum bit (qubit) resources of current quantum computers and process merely a small $2\times2$ grayscale image, far from being applicable to SRS. 
Recently, the novel concept of hybrid quantum-classical GAN, a quantum generator with a classical discriminator, has upgraded the order to $28\times28$ (still grayscale), whereas it is still insufficient for SRS.
This motivates us to design a radically new hybrid framework, where both generator and discriminator are hybrid architectures.
We demonstrate this feasibility, leading to a breakthrough of processing $128\times128$ hyperspectral images for SRS.
Specifically, we design the quantum part with mathematically provable quantum full expressibility (FE) to address core signal processing tasks, wherein the FE property allows the quantum network to realize any valid quantum operator with appropriate training.
The classical part, composed of convolutional layers, treats the read-in (compressing the optical information into limited qubits) and read-out (addressing the quantum collapse effect) procedures.
The proposed innovative hybrid quantum GAN, named ``Hyperspectral Knot-like IntelligeNt dIscrimiNator and Generator" (HyperKING), where ``knot” partly symbolizes the quantum entanglement and partly the compressed quantum domain in the central part of the network architecture.
HyperKING significantly surpasses the classical approaches in hyperspectral tensor completion, mixed noise removal (about 3dB improvement), and blind source separation results.
\end{abstract}


    \ifconfver \else \vspace{-0.0cm}\fi

    \ifconfver \else \vspace{-0.5cm}\fi

    \ifconfver \else  \fi

\section{Introduction}\label{sec:introduction}
Quantum remote sensing technology has been very recently developed for real-time hyperspectral satellite data restoration, where quantum computing has also been shown to be promising for developing high-performance satellite remote sensing (SRS) software \cite{HyperQUEEN}.
To understand the strength of quantum computing, we remark that for a specific sampling task that would take about eight years in current computing facility, the well-known ``Zuchongzhi'' quantum computer can accomplish the task in approximately one hour \cite{upama2022evolution}.
Quantum computing is also effective in addressing NP-hard problems, such as automatic financial crime detection from the transaction records \cite{camino2017finding}, fast searching via the well-known Grover's algorithm \cite{boyer1998tight}, as well as the factorization of an integer into prime factors using the Shor's algorithm \cite{shor1999polynomial}.
Beyond these quantum applications, another leading approach is quantum machine intelligence, or more specifically quantum deep network (QUEEN) \cite{HyperQUEEN,10113742}.
For instance, QUEENs significantly improve the performance of graph and convolutional neural networks by integrating the quantum unitary-computing and classical deep features in very recent SRS literature \cite{QuantumHCD,quantumMM}.
Remarkably, QUEEN-inspired quantum prism has even been adopted to generate more virtual spectrum observations in order to solve an NP-hard underdetermined blind source separation problem for optical satellite data analysis \cite{PRIME}.
Another critical SRS application is object counting \cite{Counting1}, whereas the relatively tiny shape of remotely sensed objects hinders effective and precise counting \cite{Counting2}.
In the future, this challenging task may be effectively and precisely achieved by exploring other entanglement mechanisms for the design of QUEENs.

\begin{figure}[t]
    \centering
   \includegraphics[width=1\linewidth]{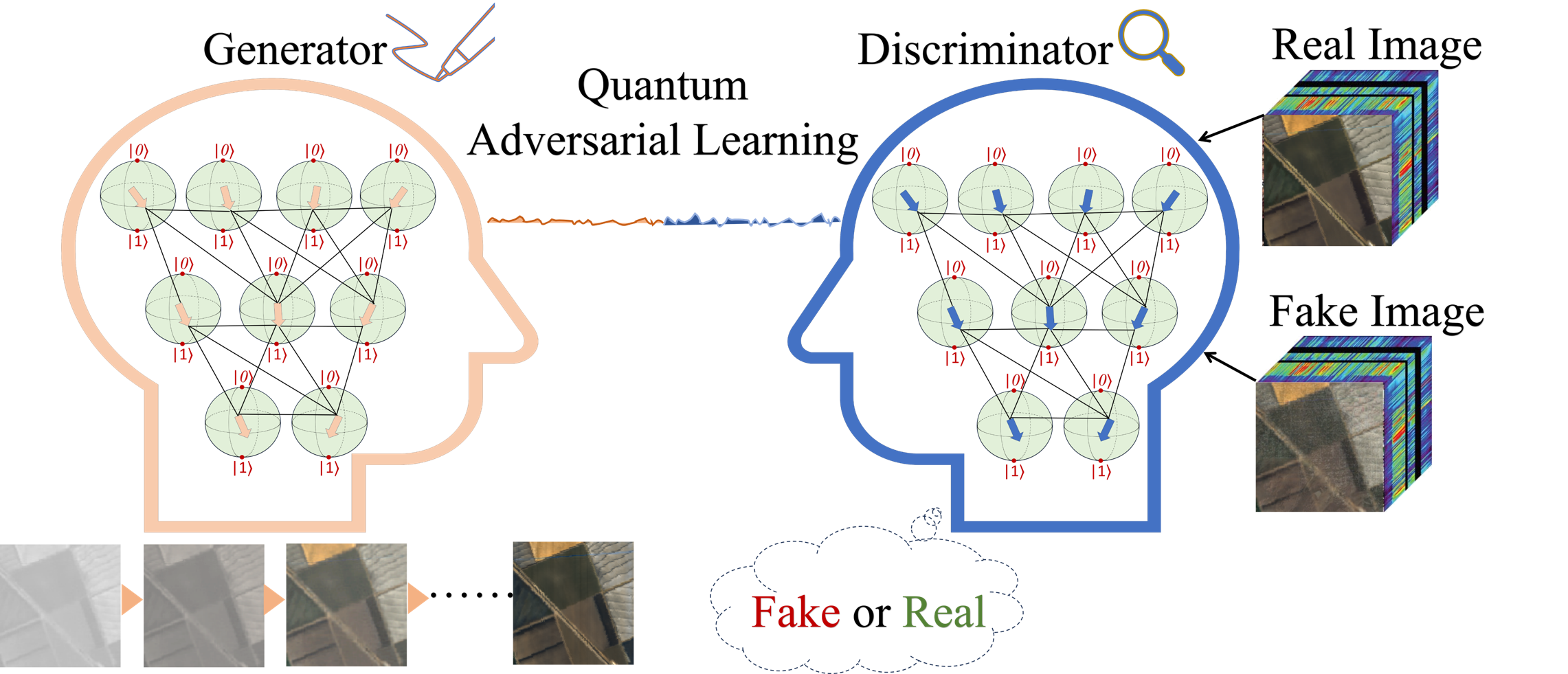}
     \vspace{-0.2cm}
    \caption{Illustration of the quantum adversarial learning via game theory, where the two players are the quantum generator and the quantum discriminator. The former aims at generating something plausible, while the latter should judge whether something is real of fake. Both the generator and discriminator have quantum logical gates as their neurons.}\label{fig:GANillus}
   \vspace{-0.3cm}
\end{figure}

Given the successes of QUEENs, we aim to advance further by developing quantum-based generative artificial intelligence (AI).
Specifically, among the various types of generative AIs (e.g., variational autoencoder or diffusion model), generative adversarial network (GAN) is regarded as the most prestigious and representative one with great successes over the past decade \cite{GAN_intro}.
Unfortunately, traditional GANs often suffer from prohibitive training datasets and hardware requirements \cite{PatchGAN}, leading to inefficient implementations.
Recently, numerous studies \cite{PatchGAN,zhou2023hybrid,PQGAN} have demonstrated that quantum-based GANs encompass great potential to outperform their classical counterparts.
This fact will also be qualitatively and quantitatively verified on the challenging mixed noise removal tasks (cf. Section \ref{subsec:Denoising}).
Consequently, constructing the quantum-based GANs has gained considerable attention and become an emerging topic in the SRS field. 
To further clarify our research aim, we first remark that quantum GAN was conceptually proved to be a feasible approach, though still far from being applicable to processing of large images.
Typically, a quantum GAN comprises two players, i.e., quantum generator and quantum discriminator.
The former aims to produce plausible data, while the latter tries to distinguish the statistical discrepancy between the real data and the fake one generated from the generator (cf. Figure \ref{fig:GANillus}).
Under the quantum adversarial learning (QAL), the dynamic game would converge to some Nash equilibrium yielding a capable quantum generator with quite accurate distributions.
Nevertheless, the existing quantum GAN could only process small-size grayscale images due to the inevitable restrictions of near-term quantum computers, and is hence not applicable to the multi-channel image processing in SRS.

To better understand the challenge, we summarize the comparison among the existing quantum-based GANs in Table \ref{QGAN_comparison}.
For instance, the pioneering QuGAN \cite{QuGAN} returns only $2\times 2$ samples per run, followed by performing inverse principal component analysis (PCA) to complete the image generation.
Subsequently, by leveraging the innovative concept of hybrid quantum-classical GAN (i.e., quantum generator with classical discriminator), the order of resolution is directly upgraded to $8\times 8$ grayscale pixels \cite{PatchGAN}.
Later on, the hybrid GAN further achieves a resolution of $28\times 28$ for grayscale image generation, as proposed recently in \cite{PQGAN}.
These all demonstrate the effectiveness of the hybrid approach, most importantly emphasizing the necessity of complementary techniques of quantum computing, which enables advanced functionalities on near-term quantum computers, as alluded in \cite{marvian2022restrictions}.
For example, deep learning has served as a complementary technique to assist quantum computing in achieving complicated quantum image processing tasks like salt-and-pepper noise removal \cite{HyperQUEEN}.

However, these hybrid GANs still fall within the scope of grayscale image processing \cite{PQGAN}, marking a deficient spectral capacity of the existing quantum-based GANs.
Whereas optical SRS, in particular, hyperspectral SRS, highly relies on the spectral capacity for accurate object identification \cite{EMI2015}.
Before we further specify the challenges of trying to
apply these approaches for SRS, we would like to ask a simple question: 
\textit{If the hybrid architecture is so effective, why not just bring the hybrid design directly into the generator (and also the discriminator)?}
This motivates us to design a radically new hybrid GAN architecture, directly making both generator and discriminator \textit{hybrid}.
In other words, existing methods all have the hybrid form with respect to (w.r.t.) the entire GAN, while each network (generator or discriminator) still remains either quantum or classical architecture, as summarized in Table \ref{QGAN_comparison}.
We aim to demonstrate that our new design philosophy is feasible, and our proposed quantum-classical GAN leads to a breakthrough of not only achieving $128 \times 128$ spatial resolution but also up to 172-channel hyperspectral SRS image processing (much upgraded compared to the single-channel grayscale image).

\begingroup
\renewcommand{\arraystretch}{1.15} 
\begin{table*}[t]\label{QGAN_comparison}
\footnotesize
\centering
    \caption{Comparison among the proposed HyperKING and existing quantum-based GANs for image processing.}
\begin{tabular}{cc||cccc||c}
    \toprule
    & Methods$~$&$ $Dataset & Image Size & Image Type & Compression & Hybrid State (Generator~\!/~\!Discriminator) 
     \rule{0pt}{2.3ex}\\
    \hline \hline
    &\multirow{2}{*}{QuGAN{\cite{QuGAN}}} & MNIST & \multirow{2}{*}{$2\times 2$} &  Grayscale &  \multirow{2}{*}{PCA Feature}   & Quantum~\!/~\!Quantum   
    \\
     & & Image & &  (single band) &    &  (Non-Hybrid)   
    \\
    \hline
    & \multirow{2}{*}{Patch GAN{\cite{PatchGAN}}} & Handwritten & \multirow{2}{*}{$8\times 8$} &  Grayscale & \multirow{2}{*}{Patch-and-Batch} & Quantum~\!/~\!Classical 
    \\
     &  &  Digits Image & &  (single band) &  &   (Hybrid)
    \\
    \hline
    & \multirow{2}{*}{PQWGAN{\cite{PQGAN}}} & Fashion & \multirow{2}{*}{$28\times 28$} & Grayscale & \multirow{2}{*}{Patch-and-Batch} & Quantum~\!/~\!Classical 
    \\
     & \ & MNIST Image &  & (single band) &  & (Hybrid) 
     \\
     \hline
       & \multirow{2}{*}{HyperKING} & NASA's EO-1 & \multirow{2}{*}{$128\times 128\times 172$} &  Hyperspectral &  \multirow{2}{*}{Deep Feature} & (Quantum-Classical)~\!/~\!(Quantum-Classical) 
       \\
       &  & Satellite Image &  & ($> \! 100$ bands) &   & (Hybrid) 
       \\
    \bottomrule
\end{tabular}
\end{table*}
\endgroup

Before we elaborate on the design of the hybrid GAN, we briefly recall the challenges of current quantum computers to clarify the practicality of the proposed approach.
First, the limited qubit resources prevent the near-term quantum computers from well processing large images.
In fact, the number of entangled qubits could be quite limited to be within 100 for current quantum computers \cite{ball2021first}.
Specifically, the well-known ``IBM Eagle" is composed of just 127 qubits, while IBM has announced another quantum computer ``IBM Osprey" with 433 qubits.
The key scientific issue here is how to build quantum mechanics allowing multiple particles to stay entangled.
On the other hand, when one attempts to read out the quantum image state from the quantum computer, it will collapse to some eigenstate, making the original quantum state not directly observable.
Under these practical constraints, this work aims to demonstrate that hundred-scale qubits would be sufficient to realize the quantum generative AI with advanced hyperspectral SRS applications.

To design our hybrid quantum-classical generator, we employ the low-rank nature of hyperspectral data to design a specific QUEEN, leading to an architecture that first processes the spatial information and then the spectral information.
Specifically, the classical part of the proposed generator plays a role in compressing the signal into the highly compressed quantum feature space, making the quantum image processing feasible under the limited qubit resources.
In comparison, the PCA compression used in existing quantum GANs may only preserve 28\% data variance for the QUEEN to generate images \cite{QuGAN}.
If we adopt this less effective compression (i.e., dividing SRS images into several grayscale patches) for SRS applications, the generated images would not preserve high-fidelity spectral characteristics, potentially leading to serious misidentification for some spectrum-driven SRS tasks, such as hyperspectral anomaly detection \cite{TGFAAD}.
Other quantum-based GANs mostly using patchwise processing strategy (cf. Table \ref{QGAN_comparison}) overlook the importance of spatial relations in processing SRS images.
Such limitation may hinder those critical SRS tasks that highly rely on spatial continuity.
For instance, in hyperspectral change detection, the changes are known to be coherently associated with their neighboring regions \cite{CODEHCD}.
Different from the abovementioned compression approaches for addressing the limited qubit resources, we adopt the classical components to effectively and efficiently compress the spectral and spatial information simultaneously; subsequently, the quantum part of our hybrid generator performs the core quantum signal processing tasks.
Remarkably, unlike the conventional deep learning module that is often considered as a blackbox with unknown capability, our hybrid generator cleverly deploys quantum neurons on the deep network to ensure the quantum full expressibility (FE) of the proposed core quantum module. 
The FE property means that for any valid quantum unitary operator, there exist network parameters such that the core quantum module exactly implements the operator.
The related theoretical guarantee of FE will be detailed in Section \ref{sec:Method}.
After the core quantum signal processing stage, the classical deep learning is adopted to learn the inverse mapping from the collapsed eigenstate back to the target quantum image state, thereby mitigating the quantum collapse (QC) effect and again emphasizing the importance of introducing the complementary techniques for implementing advanced tasks on near-term quantum computers. 
Overall, our hybrid generator has a knot-like shape, where the term ``knot" partly symbolizes the concept of quantum entanglement and partly symbolizes the highly compressed quantum feature space (wherein core quantum signal processing happens) for saving the qubit resources.
\begin{figure}[t]
    \centering
   \includegraphics[width=1\linewidth]{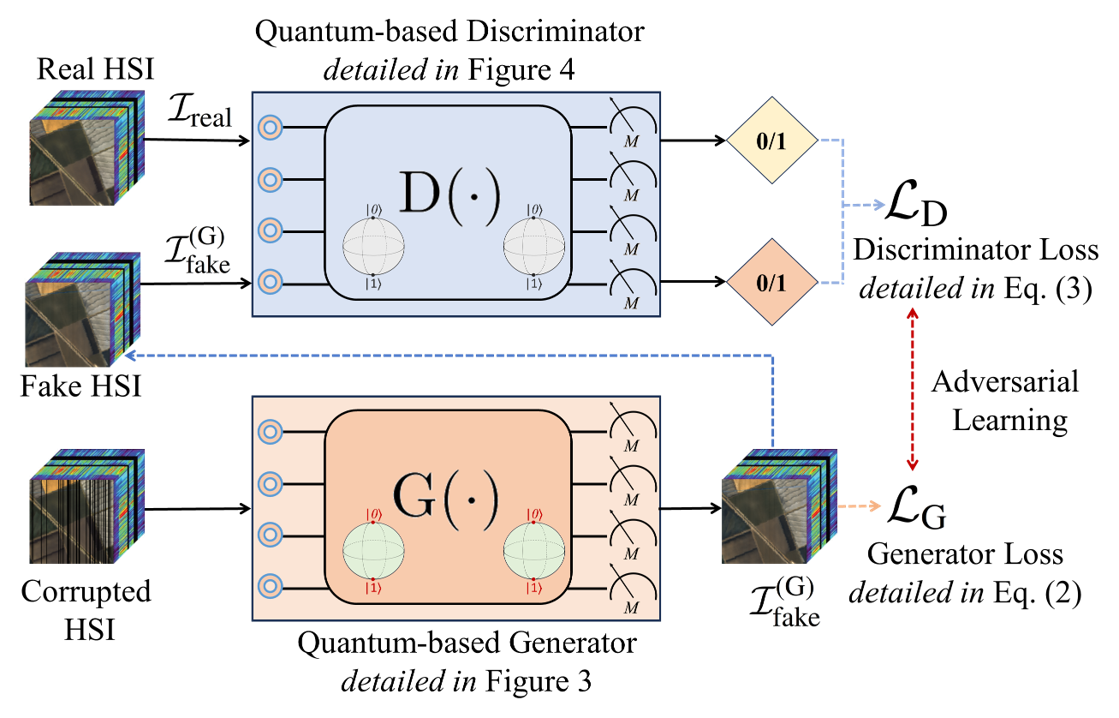}
     \vspace{-0.5cm}
    \caption{Overall framework of the proposed HyperKING, where $\text{G}(\cdot)$ and $\text{D}(\cdot)$ denote the function of the quantum-based generator and discriminator, respectively.
    Quantum adversarial learning is conducted through the adversarial loss functions between the two competitive players.}\label{fig:overall_arch}
   \vspace{-0.3cm}
\end{figure}

As for the design of our hybrid classifier, it actually relates to another challenging issue, which is to ensure that the proposed hybrid generator and discriminator can help each other to alternatively evolve to stronger statuses under the QAL.
To this end, the hybrid discriminator may be simply considered and constructed as a highly-entangled deep classifier.
Specifically, the function of the quantum-driven discriminator is regarded as to classify a hyperspectral image to be fake or real (cf. Figure \ref{fig:GANillus}), and this function should be sufficiently strong for the discriminator to adversarially help the proposed hybrid generator to be trained stronger under the dynamic game theory framework \cite{fudenberg1991game}.
The proposed hybrid discriminator is composed of a series of quantum rotation gates in a highly-entangled form (just like a knot), which is known to be an effective strategy for the classification task \cite{sim2019expressibility}.
The key challenge here is to ensure that our discriminator is neither too strong nor too weak, so that the generator and discriminator can be successfully evolved via the QAL game \cite{GAN}.
Comprehensive analyses, as summarized in Section \ref{subsec:Comparability Analysis}, prove the balanced competitivity between the proposed generator and discriminator.
Once the game converges, the hybrid generator is expected to be strong enough to be applicable to challenging tasks (e.g., hyperspectral satellite image restoration).
The proposed method is termed ``Hyperspectral Knot-like IntelligeNt dIscrimiNator and Generator" (HyperKING), which is the first quantum generative AI that achieves advanced non-grayscale quantum image processing tasks.
To better understand our advanced quantum-based generative AI, the overall architecture of HyperKING is provided in Figure \ref{fig:overall_arch}.

The remaining parts of this paper are organized as follows.
In Section \ref{sec:Method}, we propose a novel quantum-classical generative AI theory and framework, including the design of the hybrid generator, the design of hybrid discriminator, and the mathematical guarantee of quantum expressibility with mathematical details collectively provided in Section \ref{sec:proof-FE}.
In Section \ref{sec:expALL}, we apply our theory to show that quantum-driven SRS is feasible, demonstrate that the proposed hybrid generator and discriminator do have competitive ability to help each other to evolve to a stronger state, and prove that the trained hybrid generator achieves good results in hyperspectral tensor completion, mixed noise removal and blind source separation.
Concluding remarks are drawn in Section \ref{sec:Conclusion}.

\begin{table}[t]
\caption{\label{tab:common_qu_gate} Frequently used quantum gates, their symbols and the corresponding unitary operators \cite{crooks2020gates, nielsen2002quantum}, where $c_\theta\triangleq\cos(\theta/2)$,
$s_\theta\triangleq\sin(\theta/2)$, and $\textrm{DIAG}(\bY_1,\dots,\bY_N)$ denotes the block-diagonal matrix with $\bY_n$ being the $n$th diagonal block for $n=1,\dots,N$ \cite{CVXbookCLL2016}.}
\begin{center}
\begin{tabular}{|c c c|} 
 \hline
 \rule{0pt}{2ex}
 Quantum Gate & Symbol & Unitary Operator 
 \rule{0pt}{2ex}
 \\
 \hline
 \rule{0pt}{4ex}
 Rotation X
 &
 \begin{tikzcd}
    \qw & \gate{R_{X}(\theta)} & \qw
 \end{tikzcd}
 &
 $\begin{pmatrix}
    c_\theta & -i s_\theta \\
    -i s_\theta & c_\theta
\end{pmatrix}$
 \rule{0pt}{4ex}
 \\ 
 \hline
 \rule{0pt}{4ex}
 Rotation Y
 &
 \begin{tikzcd}
    \qw & \gate{R_{Y}(\theta)} & \qw 
 \end{tikzcd}
 & 
 $\begin{pmatrix}
   c_\theta & - s_\theta \\
    s_\theta & c_\theta
\end{pmatrix}$
 \rule{0pt}{4ex}
\\
\hline
 \rule{0pt}{4ex}
 Rotation Z
 &
 \begin{tikzcd}
    \qw & \gate{R_{Z}(\theta)} & \qw 
 \end{tikzcd}
 & 
 $\begin{pmatrix}
    e^{-i (\theta/2)} & 0 \\
    0 & e^{i(\theta/2)}
\end{pmatrix}$
 \rule{0pt}{4ex}
  \\
 \hline
 \rule{0pt}{6.5ex}
 Ising XX
 &
\begin{quantikz}
    \qw & \gate{XX(\theta)} & \qw
\end{quantikz}
&

$
\begin{pmatrix}
    c_\theta & 0 & 0 & \frac{s_\theta}{i}\\
    0 & c_\theta & \frac{s_\theta}{i} & 0\\
    0 & \frac{s_\theta}{i} & c_\theta & 0\\
   \frac{s_\theta}{i} & 0 & 0 & c_\theta
\end{pmatrix}
$

\\
\hline
\rule{0pt}{4ex}
 Pauli-X
 &
 \begin{tikzcd}
    \qw & \gate{X} & \qw
 \end{tikzcd}
 &
 $\begin{pmatrix}
    0 & 1 \\
    1 & 0
\end{pmatrix}$
\\
\hline
\rule{0pt}{4ex}
 Pauli-Z
 &
 \begin{tikzcd}
    \qw & \gate{Z} & \qw
 \end{tikzcd}
 &
 $\begin{pmatrix}
    1 & 0 \\
    0 & -1
\end{pmatrix}$
 \rule{0pt}{4ex}
 \\
 \hline
 \rule{0pt}{11ex}
Toffoli
 &
 \begin{tikzcd}
    \qw & \ctrl{1} & \qw \\
    \qw & \octrl{1} & \qw \\
    \qw & \targ{} & \qw
 \end{tikzcd}
 &
 $\textrm{DIAG}(\bI_4,X,\bI_2)$
 \rule[-5ex]{0pt}{4ex}
 \\
 \hline
 \rule{0pt}{7ex}
 CRX
 &
 \begin{tikzcd}
    \qw & \ctrl{1} & \qw \\
    \qw & \gate{R_{X}} & \qw
 \end{tikzcd}
 &
 $\begin{pmatrix}
    1 & 0 & 0 & 0\\
    0 & 1 & 0 & 0\\
    0 & 0 & c_\theta & -is_\theta \\
    0 & 0 & -is_\theta & c_\theta
\end{pmatrix}$
 \rule{0pt}{4ex}
 \\
 \hline
\end{tabular}
\end{center}
\end{table}

\section{The Proposed Quantum-Based Generative AI Framework}\label{sec:Method}

\begin{figure*}[t]
    \centering
   \includegraphics[width=1\linewidth]{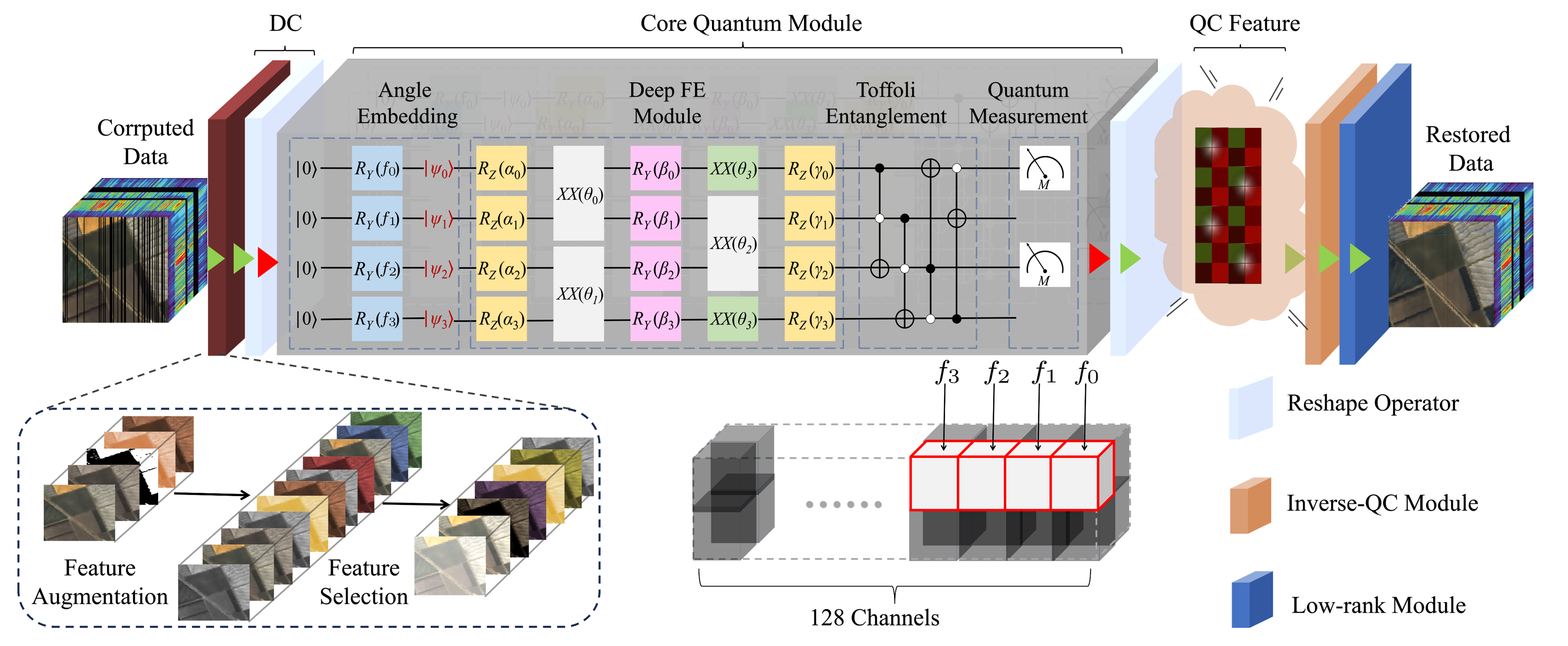}
     \vspace{-0.65cm}
    \caption{The proposed hybrid generator for hyperspectral image processing, where the QC effect and the issue of limited qubit resource are well addressed as experimentally demonstrated.
    The full expressibility of the Ising-Rotation architecture ``$R_Z-XX-R_Y-XX-R_Z$" is theoretically proved. Please refer to \cite[Figure 3]{HyperQUEEN}.}\label{fig:Gnet}
\end{figure*}

Quantum-based machine intelligence starts showing its impact on optical SRS, as featured by the recent achievement of the high-accuracy QUEEN-based satellite data restoration.
This section aims to go a step further to develop quantum-based generative AI with sufficiently large spatial capacity, in order to perform meaningful texture analysis in advanced image processing tasks.
Existing quantum GAN could only process a small $2\times 2$ grayscale image or single-channel image; after introducing the novel concept of hybrid quantum-classical GAN (cf. Table \ref{QGAN_comparison}), the resolution is upgraded to $8\times 8$ and very recently advanced to $28\times 28$ grayscale pixels.
This motivates us to consider the hybrid structure but using a radically new hybrid design, directly making both generator and discriminator in the hybrid form (cf. Table \ref{QGAN_comparison}).
Comprhensive experiments, as summarized in Section \ref{sec:expALL}, will demonstrate that our new design does lead to a breakthrough of processing $128\times 128$ SRS image with 172-channel hyperspectral bands.

In our proposed quantum-based generative AI, i.e., HyperKING, the quantum part is responsible for the core signal processing tasks with mathematically provable quantum FE, while the classical part treats the read-in (compressing the optical information into limited qubits) and read-out (addressing the QC effect) procedures, leading to the ``knot-like” architecture as discussed in Section \ref{sec:introduction}.
In HyperKING, both the highly-entangled structure of the discriminator (cf. Figure \ref{fig:Dnet}) and the highly-compressed central quantum part of the generator (cf. Figure \ref{fig:Gnet}) are symbolized by the term ``knot".

To understand the key principles behind our design, preliminary knowledge of quantum computing and quantum deep learning are required for comprehensive understanding.
We refer interested readers to \cite[Section II.B]{HyperQUEEN} and \cite[Section II.C]{HyperQUEEN}, where one can find a concise review of preliminary backgrounds, including the quantum Dirac notation system, quantum measurement, QC effect, barren plateaus effect, Bloch sphere, and quantum deep learning, etc.
The following sections describe the design philosophy of the proposed quantum-based generative AI, detailing the proposed hybrid generator and hybrid discriminator, followed by some analysis and discussion.
By illustrating our design philosophy, we aim to shed some insights for future investigators in this pioneering research line.

\subsection{HyperKING: The Design of Hybrid Generator and Hybrid Discriminator}
Generative AI has found numerous applications in various domains \cite{cooper2023examining}.
Oriented from the game theory, a game is composed of three components---the players in the game, their strategy spaces, and the utility functions \cite{fudenberg1991game}.
As a typical example of generative AI, there are two players (or networks), which are referred to as the generator and the discriminator.
The strategy space of the generator is the parameter space of the generative network, aiming at generating plausible samples that are capable of misleading the discriminator.
The strategy space of the discriminator is the parameter space of the discriminative network, aiming at discriminating whether given samples are real or fake.
Eventually, the utility of the generator is evaluated by its capability of deceiving, while the utility of the discriminator is evaluated by its resilience against being deceived.
The above idea is graphically illustrated in Figure \ref{fig:GANillus}.
As for the proposed quantum-based generative AI, quantum neurons (i.e., quantum gates) are used to deploy the hybrid generator/discriminator networks.
To the best of our knowledge, the demonstration of quantum-based generative AI for processing multi-channel, sufficiently large images (needed in real-world SRS applications) is absent.
For the first time, we design an advanced hybrid GAN framework, and apply it to advanced tasks in SRS.

To design the hybrid generator, recall that our target application is to restore hyperspectral data, meaning that the generator needs to output a hyperspectral image (HSI) that is, however, known to be of low rank.
Under the convex optimization framework, such a low rankness can be forced by suitable regularization functions like nuclear norm \cite{CVXbookCLL2016}.
However, enforcing a generative network to return a low-rank HSI, denoted as $\bX$, is relatively challenging.
This is equivalent to requiring the network itself to fulfill the mission of low-rank regularization, which is closely related to the well-known concept called deep image prior (DIP) \cite{DIP}.
Instead of designing some convex regularization/loss functions \cite{SSSS}, the DIP concept argues that the network itself actually implements some specific regularization.
Inspired by the low-rank modeling $\bX=\bA\bS$ \cite{HISUN} wherein $\bA$ encodes the spectral information, while $\bS$ encodes the spatial details, our hybrid generator first addresses the spatial details $\bS$ through a customized QUEEN, and then performs the spectral upsampling (namely, left-multiplying by $\bA$) at the last classical layer as illustrated in Figure \ref{fig:Gnet} and detailed in Table \ref{tab:configuration}, i.e., low-rank module $E_r$ \cite{HyperQUEEN}.

Besides the low-rank design, deep learning also offers a great environment for saving the limited qubit resources.
Specifically, deep learning can be regarded as to process the signal in the highly compressed feature space, rather than directly processing the original hyperspectral signal of huge data volume.
This idea of deep compression is more effective than patchwise computing or PCA compression used in existing
benchmark quantum-based GANs (cf. Table \ref{QGAN_comparison}), and is expected to capture more non-linear effects of real-world satellite data for subsequent quantum processing.

Thus, at the beginning of the proposed hybrid generator, we add a deep compression (DC) module to transform the hyperspectral information into the highly compressed feature space, wherein we conduct the core quantum information processing (i.e., the core quantum module), as illustrated in Figure \ref{fig:Gnet} and detailed in Table \ref{tab:configuration}, where ``MCB($x$,$y$)" denotes the multiple convolution block (CB) composed of $2\times 2$ max pooling, CB($x$,3,0), CB(2$x$,3,1), and CB(2$x$,3,0)$\times y$; note that CB($c$,$s$,$p$) is defined in Table \ref{tab:configuration}.
Before formally designing the core quantum module, we first discuss the QC effect, which prevents us from smoothly reading out the well processed/restored quantum image from the quantum computer.
To tackle this challenge, we adopt the inverse-QC module to learn the inverse mapping from the measured QC features back to the original data space, as illustrated in Figure \ref{fig:Gnet} and detailed in Table \ref{tab:configuration}.
With all the above keeping in mind, we are ready to design the quantum components (i.e., core quantum module), the most challenging and tricky part for the core quantum signal processing tasks.

First of all, the core quantum module adopts the quantum angle encoding scheme \cite{weigold2021encoding}, for which each feature element $f$ is encoded as the angle of the quantum rotation gate $R_Y$ (i.e., $R_Y(\theta)|_{\theta:=f}$) for the subsequent quantum information processing (cf. Figure \ref{fig:Gnet}), where $R_Y$ is defined in Table \ref{tab:common_qu_gate}.
The way we treat the quantum features are graphically illustrated in Figure \ref{fig:Gnet}.
One can see that we are not treating one channel at once, but treating $f_k$'s from different channels at once.
The reason is that the success of classical deep learning would come from the interaction among different feature maps (i.e., channels), allowing the extraction of more abstract features in the subsequent layers (i.e., deeper layers), and our way of treating the quantum feature allows more interaction among different channels.
Second, we deploy the 5-layer ``$R_Z-XX-R_Y-XX-R_Z$" architecture to process the quantum features.
Unlike most deep learning works wherein the network expressibility is often unknown (blackbox), we prove that the above Ising-Rotation architecture has full expressibility (FE), meaning that the deployed quantum neurons are able to express/implement any valid quantum unitary operator.
This is rigorously stated in the following theorem, whose proof is relegated to Section \ref{sec:proof-FE}.
\begin{center}
\scriptsize
\begin{table}[t]
    \caption{Architecture of the proposed hybrid generator.
    ``CB($c$,$s$,$p$)" denotes convolution block composed of the convolution layer ``Conv($c$,$s$,$p$)", batch normalization (BN) and LeakyReLU (with negative slope 0.2), while ``TCB($c$,$s$)" denotes transposed CB composed of transposed convolution TConv($c$,$s$), BN and LeakyReLU, where $(c,s,p)$ specify the number of output channels, the kernel size of $s\times s$, and the number of paddings, respectively.
    The Up-sampling layer is implemented by bilinear interpolation.}
    \centering
    \label{tab:configuration}
\begin{tabular}{c||c|c|c}
    \hline
    \hline
    \rule{0pt}{2.3ex}
     & Layer & Configuration & Output Size
    \rule{0pt}{2ex}
     \\
    \hline
    \hline
    \rule{0pt}{2.3ex}
    \multirow{1}{*}{} & Input & - & 172$\times$128$\times$128
    \rule{0pt}{2ex}
    \\
    
    \hline
    \rule{0pt}{2.3ex}
    \multirow{8}{*}{\makecell{DC Module}} & \multirow{4}{*}{ ConvModule 1}  & CB(516,3,1)$\times$1 & \multirow{4}{*}{16$\times$124$\times$124}
    \rule{0pt}{2.3ex}\\
     & & CB(172,3,1)$\times$1  & 
    \rule{0pt}{2.3ex}\\
     & & CB(32,3,1)$\times$1  & 
    \rule{0pt}{2.3ex}\\
     & & CB(16,3,0)$\times$2  & 
    \rule{0pt}{2.3ex}\\
    \cline{2-4} &  \multirow{1}{*}{ConvModule 2} &  MCB(16,2)$\times$1 & \multirow{1}{*}{32$\times$56$\times$56}
    \rule{0pt}{2.3ex}\\
    \cline{2-4} & \multirow{1}{*}{ConvModule 3} & MCB(32,3)$\times$1 & \multirow{1}{*}{64$\times$20$\times$20}
    \rule{0pt}{2.3ex}\\
    \cline{2-4} & \multirow{1}{*}{ConvModule 4} & MCB(64,3)$\times$1 & \multirow{1}{*}{128$\times$2$\times$2}
    \rule{0pt}{2.3ex}\\
    \hline
    \multirow{21}{*}{\makecell{Core\\Quantum\\FE Module\\(Theorem 1)}} & Reshape  & - & 128$\times$4
    \rule{0pt}{2.3ex}\\
    \cline{2-4} & \makecell{Data\\Embedding} & $R_{Y}(128,4)$ & 128$\times$4
    \rule{0pt}{2.5ex}\\
    \cline{2-4} & Unitary Gate 1 & $R_{Z}(128,4)$ & 128$\times$4
    \rule{0pt}{2.3ex}\\
    
    \cline{2-4} & Reshape  & - & 256$\times$2
    \rule{0pt}{2.3ex}\\
    \cline{2-4} & Unitary Gate 2 & $XX(256,2)$ & 256$\times$2
    \rule{0pt}{2.3ex}\\
    
    \cline{2-4} & Reshape  & - & 128$\times$4
    \rule{0pt}{2.3ex}\\
    \cline{2-4} & Unitary Gate 3 & $R_{Y}(128,4)$ & 128$\times$4
    \rule{0pt}{2.3ex}\\
    
    \cline{2-4} & Reshape  & - & 256$\times$2
    \rule{0pt}{2.3ex}\\
    \cline{2-4} & Unitary Gate 4 & $XX(256,2)$ & 256$\times$2
    \rule{0pt}{2.3ex}\\
    
    \cline{2-4} & Reshape  & - & 128$\times$4
    \rule{0pt}{2.3ex}\\
    \cline{2-4} & Unitary Gate 5 & $R_{Z}(128,4)$ & 128$\times$4
    \rule{0pt}{2.3ex}\\
    \cline{2-4} & \multirow{4}{*}{\makecell{Toffoli\\Entanglement}} & CCNOT(0,1,2) & 128$\times$4
    \rule{0pt}{2.3ex}\\
    \cline{3-4} &  & CCNOT(1,2,3) & 128$\times$4
    \rule{0pt}{2.3ex}\\
    \cline{3-4} &  & CCNOT(2,3,0) & 128$\times$4
    \rule{0pt}{2.3ex}\\
    \cline{3-4} &  & CCNOT(3,0,1) & 128$\times$4
    \rule{0pt}{2.3ex}\\
    \cline{2-4} & \makecell{QC\\Measurement} & $Z(128,2)$ & 128$\times$2
    \rule{0pt}{3ex}\\
    \cline{2-4} & Reshape  & - & 64$\times$2$\times$2
    \rule{0pt}{2.3ex}\\
    
    \hline
    \multirow{12}{*}{\makecell{Inverse-QC\\Module}} & \multirow{2}{*}{TConvModule 1} & TCB(64,3)$\times$4 & \multirow{2}{*}{64$\times$20$\times$20}
    \rule{0pt}{2.3ex}\\
    & & 2$\times$2 Up-sample &
    \rule{0pt}{2.3ex}\\
    \cline{2-4} & \multirow{3}{*}{TConvModule 2} & TCB(64,3)$\times$2 & \multirow{3}{*}{32$\times$56$\times$56}
    \rule{0pt}{2.3ex}\\
     &  & TCB(32,3)$\times$2 & 
    \rule{0pt}{2.3ex}\\
    & & 2$\times$2 Up-sample &
    \rule{0pt}{2.3ex}\\
    \cline{2-4} & \multirow{3}{*}{TConvModule 3} & TCB(32,3)$\times$2 & \multirow{3}{*}{16$\times$124$\times$124}
    \rule{0pt}{2.3ex}\\
     &  & TCB(16,3)$\times$1 & 
    \rule{0pt}{2.3ex}\\
    & & 2$\times$2 Up-sample &
    \rule{0pt}{2.3ex}\\
    \cline{2-4} & \multirow{2}{*}{TConvModule 4} & TCB(16,3)$\times$1 & \multirow{2}{*}{8$\times$128$\times$128}
    \rule{0pt}{2.3ex}\\
     &  & TCB(8,3)$\times$1 & 
    \rule{0pt}{2.3ex}\\
    \hline
    \multirow{2}{*}{\makecell{Low-rank\\Module}} & \multirow{2}{*}{$E_r$ Module} & CB(172,1,0)$\times$1 & \multirow{2}{*}{172$\times$128$\times$128}
    \rule{0pt}{2.3ex}\\
     & & Conv(172,1,0)$\times$1  &
    \rule{0pt}{2.3ex}\\
    \hline
\end{tabular}
\end{table}
\end{center}

\begin{Theorem}\label{thm:FE-YXIsing}
The learnable quantum neurons deployed in the core quantum module of the proposed hybrid generator (cf. Figure \ref{fig:Gnet}) can realize any valid quantum operator $U$, with some real-valued network parameters $\{\alpha_{k},\beta_{k},\gamma_{k},\theta_{k}\}$.
\hfill$\square$
\end{Theorem}

\noindent
Third, the core quantum module adopts the Toffoli entanglement, implemented by the controlled CNOT (CCNOT) with the controlled state $\ket{10}$, followed by the Pauli-Z quantum measurement (cf. Table \ref{tab:common_qu_gate}) \cite{HyperQUEEN}.
Finally, to see the insight from Theorem \ref{thm:FE-YXIsing}, we remark that from an intuitive ground, being able to express comprehensive quantum operators (i.e., FE) has positive impact on ``generating" more plausible images---a desired property for GAN.

\begin{figure*}[t]
    \centering
   \includegraphics[width=1\linewidth]{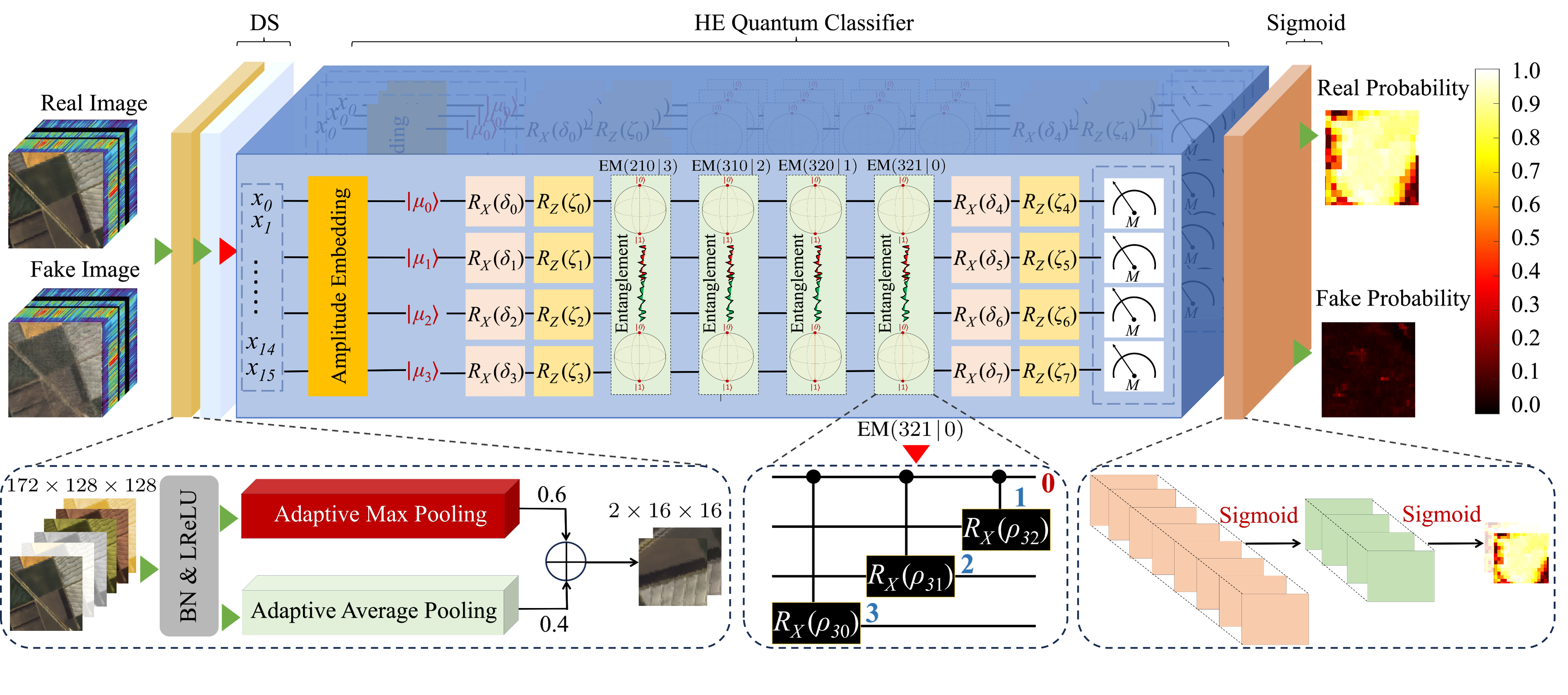}
     \vspace{-0.65cm}
    \caption{The proposed hybrid discriminator for highly-entangled (HE) quantum classification, where the HE property is achieved by the entanglement module $\textrm{EM}(bcd~\!\!\!\mid~\!\!\!a)$ meaning that qubit $a$ controls qubits $(b,c,d)$ via the controlled rotation X (CRX) gates.}\label{fig:Dnet}
\end{figure*}

As for the design of our hybrid discriminator, we strategically reformulate its function as a classifier that determines whether a given hyperspectral image is real or generated.
This allows us to judiciously design the discriminator as a quantum-based classifier.
The most challenging part would be how to ensure that the designed discriminator has a comparable ability to the proposed hybrid generator; otherwise, the adversarial learning cannot be effectively proceeded, making the trained generator not strong enough to tackle the challenging satellite data restoration mission in SRS.
For this aim, we propose the following architecture for our hybrid discriminator, which is graphically illustrated in Figure \ref{fig:Dnet}.

First, before the quantum amplitude encoding \cite{QIPbook}, we propose the downsampling (DS) module that takes merit of the max pooling to have a more concise image representation for saving the qubit resources, while keeping the most significant local information.
However, considering that accurate classification would rely on some details removed by the max pooling, we mitigate the concern by further keeping those information via adding the average pooling, as displayed in Figure \ref{fig:Dnet} and detailed in Table \ref{tab:configurationD}.
Second, the core of the hybrid discriminator is designed as a highly-entangled (HE) quantum classifier.
To this end, we define the entanglement module $\textrm{EM}(bcd~\!\!\!\mid~\!\!\!a)$ that uses qubit $a$ to sequentially control the other qubits $(b,c,d)$ through the controlled rotation X (CRX) quantum gates (cf. Figure \ref{fig:Dnet}), where CRX has been defined in Table \ref{tab:common_qu_gate}.
We use consecutive EM modules to achieve the desired HE property, and this HE strategy is known to be effective as investigated in \cite{sim2019expressibility}.
Third, the quantum measurement is conducted by the Pauli-X quantum gates, followed by the sigmoid layer that transforms measured quantum feature maps into probability maps, from which the hybrid discriminator determines whether the input hyperspectral image is real or generated. 
The detailed structure of the proposed hybrid discriminator is given in Table \ref{tab:configurationD}.
Comprehensive experiments, as summarized in Section \ref{subsec:Comparability Analysis}, will evaluate the competitiveness of this elegant design compared to our hybrid generator, demonstrating the effectiveness in training the hybrid discriminator and hybrid generator through the QAL.

Our experiments also provide an insight for the design of QAL because the capability of our hybrid generator (trained with the QAL) in various SRS restoration tasks is never a coincidence.
The experiments in Section \ref{sec:expALL} will show that when the hybrid discriminator is revised to have more sophisticated or more simplified quantum structures, the QAL may not be smoothly done.
Thus, striking a balance between the generator and discriminator is one of our key contributions, allowing the two players to dynamically help each other to get stronger in the quantum adversarial game.

We call our quantum generative AI framework as ``Hyperspectral Knot-like IntelligeNt dIscrimiNator and Generator" (HyperKING)!
The term ``knot" is partly to symbolize quantum stuffs that are entangled together (e.g., the Toffoli-entangled qubits in the generator, or the CRX-entangled qubits in the discriminator), and is partly to use the compressed central part of a knot (or bowknot) to symbolize the proposed highly compressed quantum feature space for saving the qubit resources.
HyperKING is implemented on Pytorch and PennyLane platforms (the leading tool for programming quantum computers) \cite{bergholm2018pennylane, paszke2019pytorch}, and is trained using the well-known smoothed $\ell_1$-norm-aided entropy loss function \cite{ADMM-Adam} (defined in Section \ref{sec:expALL}) with the root mean square propagation (RMSProp) optimizer for addressing the non-stationary adversarial learning \cite{tieleman2012lecture}.
The superiority of the proposed HyperKING will be demonstrated in Section \ref{sec:expALL}.

\begin{center}
\scriptsize
\begin{table}[t]
    \caption{Configuration of our hybrid discriminator (cf. Figure \ref{fig:Dnet}).
    ``LReLU($m$)" denotes LeakyReLU with negative slope $m$.
    ``APool($x$,$y$,$z$)" denotes adaptive pooling module composed of adaptive max pooling ``AdaptiveMaxPool($x$,$y$,$z$)" and adaptive average pooling ``AdaptiveAvgPool($x$,$y$,$z$)", outputting a $x$-channel tensor of spatial size $y\times z$.
    Quantum amplitude embedding ``Amp($x$,$y$)" encodes $x$ features of dimension $y$ into $x$ qubit states of dimension $\log_2(y)$.
    The module ``Sigm($m$,$n$)" is composed of a trainable $m\times n$ matrix for linear projection, followed by a sigmoid function.}  
    \centering
    \label{tab:configurationD}
\begin{tabular}{c||c|c|c}
    \hline
    \hline
    \rule{0pt}{2.3ex}
     & Layer & Configuration & Output Size
    \rule{0pt}{2ex}
     \\
    \hline
    \hline
    \rule{0pt}{2.3ex}
    \multirow{1}{*}{} & Input & - & 172$\times$128$\times$128
    \rule{0pt}{2ex}
    \\
    
    \hline
    \rule{0pt}{2.3ex}
    \multirow{3}{*}{\makecell{DS Module}} & \multirow{3}{*}{\makecell{Pooling \\ Layer}}  & BN$\times$1 & \multirow{3}{*}{2$\times$16$\times$16}
    \rule{0pt}{2.3ex}\\
     & & LReLU(0.2)$\times$1  & 
    \rule{0pt}{2.3ex}\\
     & & APool(2,16,16)$\times$1  & 
    \rule{0pt}{2.3ex}\\

    \hline
    \multirow{15}{*}{\makecell{HE Quantum\\Classifier\\Module}} & Reshape  & - & 32$\times$16
    \rule{0pt}{2.3ex}\\
    \cline{2-4} & \makecell{Data\\Embedding} & {Amp}(32,16) & 32$\times$4
    \rule{0pt}{2.5ex}\\
    \cline{2-4} & Unitary Gate 1 & $R_{X}(32,4)$ & 32$\times$4
    \rule{0pt}{2.3ex}\\
    \cline{2-4} & Unitary Gate 2 & $R_{Z}(32,4)$ & 32$\times$4
    \rule{0pt}{2.3ex}\\
    
    \cline{2-4} & \multirow{4}{*}{\makecell{CRX\\Entanglement}} & $\textrm{EM}(210~\!\!\!\mid~\!\!\!3)$ & 32$\times$4
    \rule{0pt}{2.3ex}\\
    \cline{3-4} &  & $\textrm{EM}(310~\!\!\!\mid~\!\!\!2)$ & 32$\times$4
    \rule{0pt}{2.3ex}\\
    \cline{3-4} &  & $\textrm{EM}(320~\!\!\!\mid~\!\!\!1)$ & 32$\times$4
    \rule{0pt}{2.3ex}\\
     \cline{3-4} &  & $\textrm{EM}(321~\!\!\!\mid~\!\!\!0)$ & 32$\times$4
    \rule{0pt}{2.3ex}\\
    \cline{2-4} & Unitary Gate 3 & $R_{X}(32,4)$ & 32$\times$4
    \rule{0pt}{2.3ex}\\
    \cline{2-4} & Unitary Gate 4 & $R_{Z}(32,4)$ & 32$\times$4
    \rule{0pt}{2.3ex}\\
    \cline{2-4} & \makecell{QC\\Measurement} & $X(32,4)$ & 32$\times$4
    \rule{0pt}{3ex}\\
    \cline{2-4} & Reshape  & - & 1$\times$128
    \rule{0pt}{2.3ex}\\
    
    \hline
    \multirow{2}{*}{\makecell{{ Sigmoid}\\{ Module}}} & \multirow{2}{*}{\makecell{Sigmoid\\ Layer}} & Sigm(128,16)$\times$1 & \multirow{2}{*}{1$\times$1}
    \rule{0pt}{2.3ex}\\
    & & Sigm(16,1)$\times$1 &
    \rule{0pt}{2.3ex}\\ 
    \hline
\end{tabular}
\end{table}
\end{center}

\subsection{Proof of Theorem \ref{thm:FE-YXIsing}}\label{sec:proof-FE}
Before we formally detail the proof of Theorem \ref{thm:FE-YXIsing}, we provide an intuitive explanation for each step to enhance the understanding of readers unfamiliar with quantum computing.
In the \textit{Step 1}, based on the basic properties of quantum computing summarized in \cite[Section II.A]{HyperQUEEN} and \cite[Section II.B]{HyperQUEEN}, we first define the general representation of the core quantum module $U$ with specific real-valued network parameters $\{\alpha_{k},\beta_{k},\gamma_{k},\theta_{k}\}$.
Then, we only need to show that for any unitary matrix $V$ (representing a specific valid quantum function), it can be realized by the core quantum module $U$ through adapting its parameters, or more precisely, training its parameters. 
To this end, we require the key properties established in the \textit{Step 2}.
The first property, as formulated in \eqref{eq:VisZYZ}, demonstrates that $V$ can be decomposed into the $R_Z-R_Y-R_Z$ architecture (i.e., FE property) without involving quantum entanglements under a single-qubit case.
However, to accomplish the challenging SRS restoration task, the network may require powerful quantum entanglements to facilitate a specific task.
To tackle it, we further deploy the trainable entanglement gate, Ising XX, in the core quantum module, where Ising gate may be considered as a soft entanglement not really forcing the entanglement but forcing cross-qubit feature interactions.
Specifically, as to be presented in the \textit{Step 2}, the Ising gate could occasionally become transparent as $\theta_k:=0$, for which the quantum module reduces back to the above naive FE form.
This implies that the network can adaptively conduct the entanglement-like interactions as needed to achieve effective restoration.
In the \textit{Step 3}, we extend the single-qubit case to the multiple-qubit case by elegantly adopting some properties of the Kronecker product, and hence completing the proof of Theorem \ref{thm:FE-YXIsing}.
With the above intuition and outline in mind, we are ready to mathematically describe the proof of Theorem \ref{thm:FE-YXIsing} in the following steps.

\textit{Step 1:}
All the quantum gates used below have been defined in Table \ref{tab:common_qu_gate}.
To facilitate the proof, we further define the notation $XX_{mn}(\theta)$, which means that the Ising XX gate $XX(\theta)$ is operated on the $m$th and $n$th qubits.
Accordingly, let $I_{0}(\theta_{2}, \theta_{3}) \triangleq  XX_{12}(\theta_{2})\otimes XX_{03}(\theta_{3})$ and 
$I_{1}(\theta_{0}, \theta_{1}) \triangleq  XX_{01}(\theta_{0})\otimes XX_{23}(\theta_{1})$ represent the overall quantum effects of the two Ising layers in our hybrid generator.
Similarly, let $A \triangleq \bigotimes_{k = 0}^{3}e^{i p_{k}} R_{Z}(\alpha_{k})$, 
$B \triangleq \bigotimes_{k=0}^{3}R_{Y}(\beta_{k})$ and 
$C \triangleq \bigotimes_{k=0}^{3}R_{Z}(\gamma_{k})$ represent the overall quantum effects of the three rotation layers in our hybrid generator.
Furthermore, any valid quantum operator must be unitary (cf. Table \ref{tab:common_qu_gate}), because a quantum gate is built from certain Hamiltonian for a specific time, yielding a unitary time evolution based on the Schrödinger equation \cite{nielsen2002quantum}.
Thus, by letting $U_{k}$ be any unitary operator acting on the $k$th qubit for $k \in \{0,1,2,3\}$, it suffices to show that there exist real-valued network parameters $\{\alpha_{k},\beta_{k},\gamma_{k},\theta_{k}\}$ such that 
\begin{equation*}
    \begin{aligned}
    U
    \triangleq \bigotimes_{k = 0}^{3} U_{k}
    = A I_{0}(\theta_{2}, \theta_{3}) B I_{1}(\theta_{0}, \theta_{1}) C,
    \end{aligned}
\end{equation*}
and this will be proven true (thereby completing the proof of Theorem \ref{thm:FE-YXIsing}) in the following steps.

\bigskip 

\textit{Step 2:}
For any unitary matrix $V$, there exist real-valued phase $p$ and parameters $\alpha, \beta, \gamma$, such that
\begin{equation}\label{eq:VisZYZ}
V = e^{i p}R_{Z}(\alpha)R_{Y}(\beta)R_{Z}(\gamma),
\end{equation}
as stated and proved in \cite[Theorem 4.1]{nielsen2002quantum}.
Moreover, by the definition of Ising XX in Table \ref{tab:common_qu_gate} and the definition of $XX_{mn}(\theta)$ in Step 1, the setting of the real-valued parameter $\theta_k:=0$ yields
\begin{equation*}
    \begin{aligned}
    I_{0}(\theta_2,\theta_3) 
    = 
    XX_{12}(0)\otimes XX_{03}(0)
    =
    (\bI_2\otimes\bI_2)\otimes(\bI_2\otimes\bI_2),
    \\
    I_{1}(\theta_0,\theta_1) 
    = 
    XX_{01}(0)\otimes XX_{23}(0)
    =
    (\bI_2\otimes\bI_2)\otimes(\bI_2\otimes\bI_2),
    \end{aligned}
\end{equation*}
implying that we just need to show that there exist real-valued $(\alpha_k,\beta_k,\gamma_k)$ such that $U=ABC$ (cf. Step 1).

\bigskip 

\textit{Step 3:}
According to \eqref{eq:VisZYZ}, we can find real-valued $(\alpha_k,\beta_k,\gamma_k)$ to achieve the following unitary matrix decomposition, i.e.,
\[
U_k
=
e^{i p_{k}} R_{Z}(\alpha_{k}) R_{Y}(\beta_{k}) R_{Z}(\gamma_{k}),~k=0,\dots,3.
\]
Then, we recall that for matrices $\bM_1,\dots,\bM_4$ of proper dimensionality, the Kronecker product satisfies $(\bM_1\otimes\bM_2)(\bM_3\otimes\bM_4)=(\bM_1\bM_3)\otimes(\bM_2\bM_4)$.
This property of Kronecker matrix equation, together with the matrices $(A,B,C)$ defined in Step 1, gives the key relation of
\begin{equation*}
    \begin{aligned}
    ABC &= \bigotimes_{k = 0}^{3}\left[e^{i p_{k}} R_{Z}(\alpha_{k}) R_{Y}(\beta_{k}) R_{Z}(\gamma_{k})\right],
    \end{aligned}
\end{equation*}
or, equivalently, $U:=\bigotimes_{k = 0}^{3} U_k=ABC$, which completes the proof of Theorem \ref{thm:FE-YXIsing}.
\hfill$\blacksquare$

\begin{figure}[t]
    \centering
   \includegraphics[width=1\linewidth]{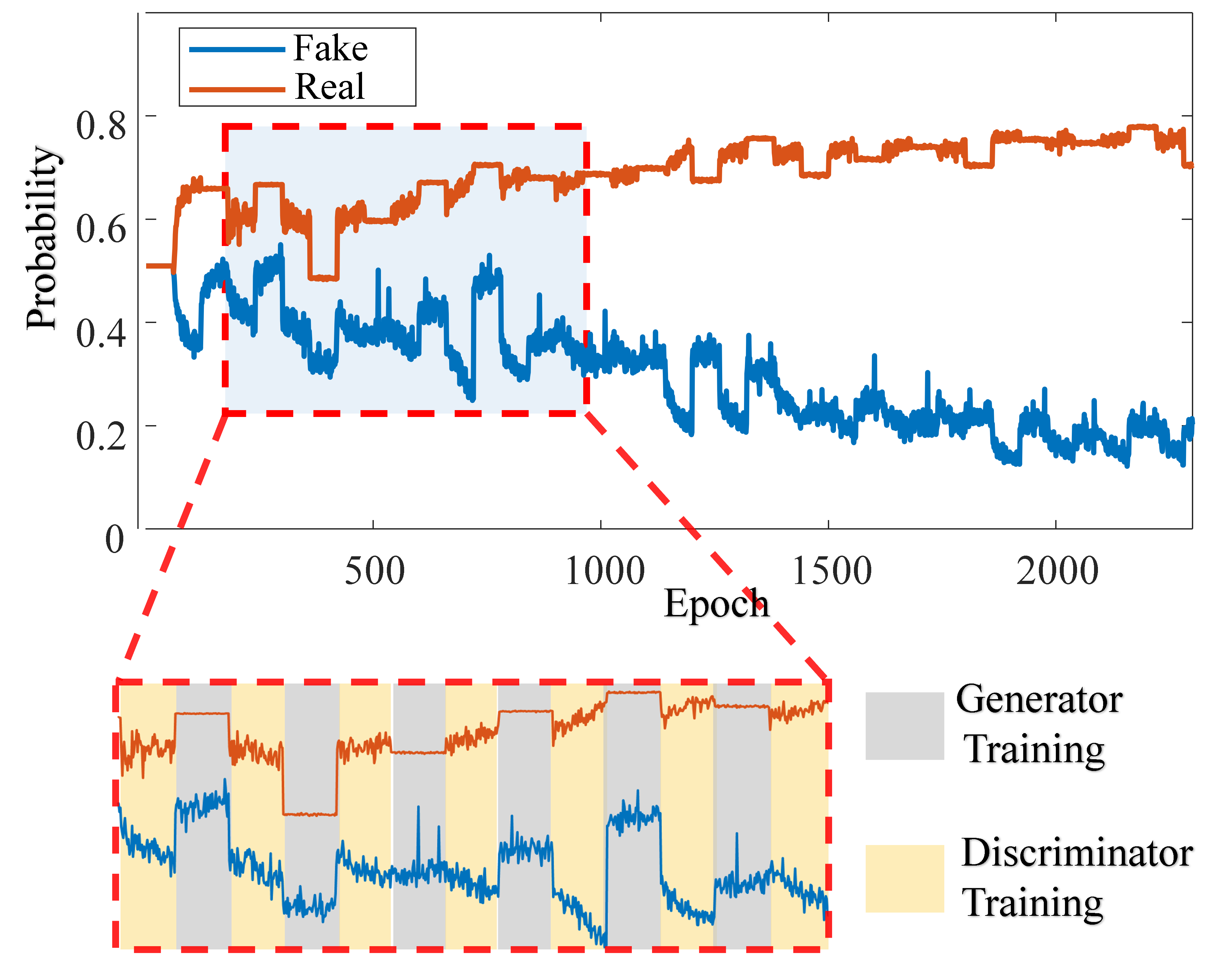}
    \caption{For a given hybrid discriminator at a specific time/epoch, the probability for the discriminator to judge a real image (resp., fake image) to be real is displayed as the orange curve (resp., blue curve), and the results are averaged over 480 independent runs.
    The gray and yellow blocks represent the training periods for the hybrid generator and discriminator, respectively.}\label{fig:QAL curve}
\end{figure}

\section{Experimental Results}\label{sec:expALL}
In this section, we comprehensively evaluate the performance of the proposed HyperKING quantum-based generative AI in the advanced SRS task (i.e., 172-channel hyperspectral image restoration).
Before we introduce the baseline methods, we remark that existing quantum-based GANs still fall within the scope of grayscale image processing (single-channel), as discussed in Section \ref{sec:introduction}.
Hence, they are not considered as baselines.
Though the hyperspectral-driven QUEEN (HyperQUEEN) \cite{HyperQUEEN} is not a quantum-based GAN (not a quantum generative AI), HyperQUEEN does achieve advanced quantum image processing (AQIP).
So, we employ the pretrained HyperQUEEN as a quantum-based baseline method.
Beyond the quantum-based architecture, we also leverage the classical baselines as there are rather limited AQIP algorithms.
Some of them are optimization-based methods, including the low-rank tensor decomposition with total variation (LRTDTV) \cite{8233403} and high-order tensor completion based on nuclear-norm and fast Fourier transform (HTNN-FFT) \cite{9730793}.
On the other hand, we also employ two DIP-based deep learning methods, i.e., hyperspectral deep image prior (HDIP) \cite{9022040} and robust deep low-rank hyperspectral inpainting (R-DLRHyIn) \cite{10032531}.

The remaining parts of this section are organized as follows.
The description of experimental settings, including the brief introduction to the baselines, and detailed description of the training/testing data, are presented in Section \ref{subsec:Experimental Settings}.
Before we investigate the performance of the proposed HyperKING, an elaborate analysis of the comparability between the hybrid generator and the hybrid discriminator is presented in Section \ref{subsec:Comparability Analysis}.
In Section \ref{subsec:Qualitative Evaluation}, Section \ref{subsec:Quantitative Evaluation}, and Section \ref{subsec:Evidencing Feasibility} we perform the qualitative evaluation, quantitative evaluation, and blind source separation [i.e., hyperspectral unmixing (HU)] evaluation, respectively, to investigate the strength of the proposed HyperKING.
Finally, we conduct hyperspectral denoising evaluation in Section \ref{subsec:Denoising} to further substantiate the effectiveness of our HyperKING.

\begin{figure}[t]
    \centering
   \includegraphics[width=0.9\linewidth]{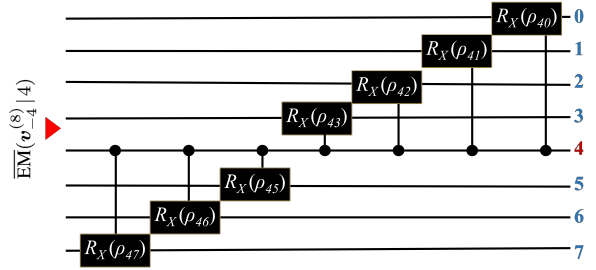}
    \caption{The schematic diagram of $\overline{\textrm{EM}}(\bv_{-4}^{(8)}~\!\!\!\mid~\!\!\!4)$ embedded within the sophisticated hybrid discriminator.}\label{fig:8qubit_circut_enta}
\end{figure}

\vspace{-0.1cm}
\subsection{Experimental Settings}\label{subsec:Experimental Settings}
Let us concisely introduce the benchmarks hereinafter.
The pioneering AQIP architecture, i.e., HyperQUEEN, is the first quantum framework that solves the image restoration problem \cite{HyperQUEEN}, but there is no quantum-driven discriminator to force HyperQUEEN to surge to a new level of AQIP performance.
Another distinction between HyperKING and HyperQUEEN is the architecture of their core quantum module.
Specifically, the generator in HyperKING is developed by the ``$R_Z-XX-R_Y-XX-R_Z$" architecture (cf. Figure \ref{fig:Gnet}), whereas HyperQUEEN comprises a ``$R_Y-XX-R_X-XX-R_Y$" design \cite{HyperQUEEN}.
This ``$R_Z-XX-R_Y-XX-R_Z$" architecture is experimentally found to enable our generator to achieve a more balanced comparability with the proposed highly-entangled quantum-driven discriminator.
As will be seen, the trained generator in HyperKING significantly outperforms other non-generative quantum or classical AIs, demonstrating the core advantage of HyperKING compared to HyperQUEEN.
To further validate this fact, we replace the original generator of HyperKING (i.e., the architecture proposed in Figure \ref{fig:Gnet}) with HyperQUEEN (i.e., the architecture proposed in \cite{HyperQUEEN}) while keeping the same training settings (to be detailed later), and we term this approach ``HyperQUEEN-QAL''.
The experiments show that the HyperQUEEN-QAL achieves an improvement by approximately 0.6dB of PSNR over HyperQUEEN;  HyperKING further surpasses HyperQUEEN-QAL by around 2dB in PSNR, substantiating the effectiveness of the ``$R_Z-XX-R_Y-XX-R_Z$" architecture and QAL.

Another benchmark HDIP leverages the deep network itself as a specific regularization scheme to take advantage of the deep image network regularization, and extend this idea to the hyperspectral domain \cite{9022040}.
Extending from HDIP, the R-DLRHyIn method not merely performs the deep prior but also takes sparse outliers into account, accordingly developing an innovative robust loss function for the low-rank restoration network \cite{10032531}.
HTNN-FFT develops a convex low-rank tensor model for high-order tensor completion based on tensor singular value decomposition (t-SVD) and Tucker nuclear norm (TNN) \cite{9730793}.
As for LRTDTV \cite{8233403}, the global spectral correlation is preserved by low-rank tensor Tucker decomposition, and an anisotropic total-variation regularization is incorporated to further characterize the image smoothness.

In this experiment, we employ a small training dataset that contains only 480 images with spatial size of 128$\times$128 pixels collected from NASA’s AVIRIS sensor \cite{AVIRISrealdata}.
Besides, we remove those water-vapor absorption spectral bands from original images \cite{HyperQUEEN}, followed by using the remaining 172 bands as reference images to obtain the training and testing data. 
To further exhibit the feasibility of our quantum-based generative AI in the SRS area, we also perform the restoration task on the real-world NASA's Earth Observation-1 (EO-1) satellite image.

As for the training settings of the proposed HyperKING, the batch size and the number of training epochs are set to 16 and 2300, respectively.
The loss functions of the hybrid discriminator and hybrid generator are presented in Section \ref{subsec:Comparability Analysis}.
Besides, we suggest the following training strategy for the proposed HyperKING to have better adversarial learning procedure.
We train the hybrid generator first, and then alternatively train the hybrid discriminator and hybrid generator with the alternating period set to 60.
Also, during the HyperKING training stage, the nonstationary discriminator loss may lead to unstable training procedure \cite{arjovsky2017wasserstein}, especially when we utilize a momentum-based optimizer (e.g., Adam \cite{kingma2014adam}).
Therefore, we employ the RMSProp optimizer \cite{tieleman2012lecture} with a learning rate of 0.01, which has conducted a promising capability in those highly non-stationary optimization problems \cite{mnih2016asynchronous}.

To understand the applicability of the proposed hybrid quantum-classical framework, we also discuss several predominant quantum computing environments, thereby providing insight of our programming strategy for future researchers.
Initially, as quantum devices have yet to be commercialized, the various quantum libraries, such as PennyLane \cite{bergholm2018pennylane}, Qiskit \cite{Qiskit}, Torch Quantum Library \cite{TorchQ}, and Tensorflow-quantum\cite{TensorflowQ}, are used to facilitate the quantum-based algorithms.
Among these platforms, PennyLane and Qiskit would be regarded as the most representative approaches for quantum programming, where PennyLane excels for those automatically differentiable algorithms; Qiskit, developed by IBM, facilitates seamless integration for the open quantum cloud resources (e.g., IBMQ \cite{IBMQ}). 
However, some literature \cite{10821196} has experimentally demonstrated that PennyLane significantly outperforms Qiskit in computational efficiency.
Furthermore, PennyLane’s lightning suite, PennyLane Lightning, provides efficient simulator to integrate the GPUs built by NVIDIA or AMD, leading to the number of qubits on a single quantum node up to 30, or even 40.
Numerous training-based quantum algorithms \cite{10881096,kwak2021introduction,PQGAN} also adopt PennyLane for quantum programming.
Thus, we consider PennyLane as the quantum programming platform for implementing and training the proposed HyperKING on large-scale data (such as hyperspectral images) to enhance the efficiency; more details of software environments are summarized in the following paragraph.
As for the hardware deployment, since our framework is a pretrained approach, it allows users to train the HyperKING through PennyLane, followed by deploying the trained/optimized quantum gate parameters on the physical quantum device (e.g., IBMQ \cite{IBMQ}).
The details of the integration of quantum software and hardware can be found in \cite{10821196}.
In summary, the proposed HyperKING is highly suitable to quantum environments, even on the larger-scale data inference.

The software environments are Python 3.10.9, PennyLane 0.25.1, Pytorch 2.1.1, and Matlab R2023a.
We leverage computational equipment composed of NVIDIA RTX 3090 GPU, AMD Ryzen 5950X CPU with 3.40-GHz speed, and 24-GB RAM to train the proposed HyperKING, and all the methods are tested on this equipment.
The comparability analysis in Section \ref{subsec:Comparability Analysis} is then conducted on the aforementioned computational equipment (for HyperKING) as well as the additional equipments composed of NVIDIA RTX 2080Ti GPU, Intel i9-10900X CPU with 3.70-GHz speed, and 12-GB RAM.

\begin{figure}[t]
    \centering
   \includegraphics[width=1\linewidth]{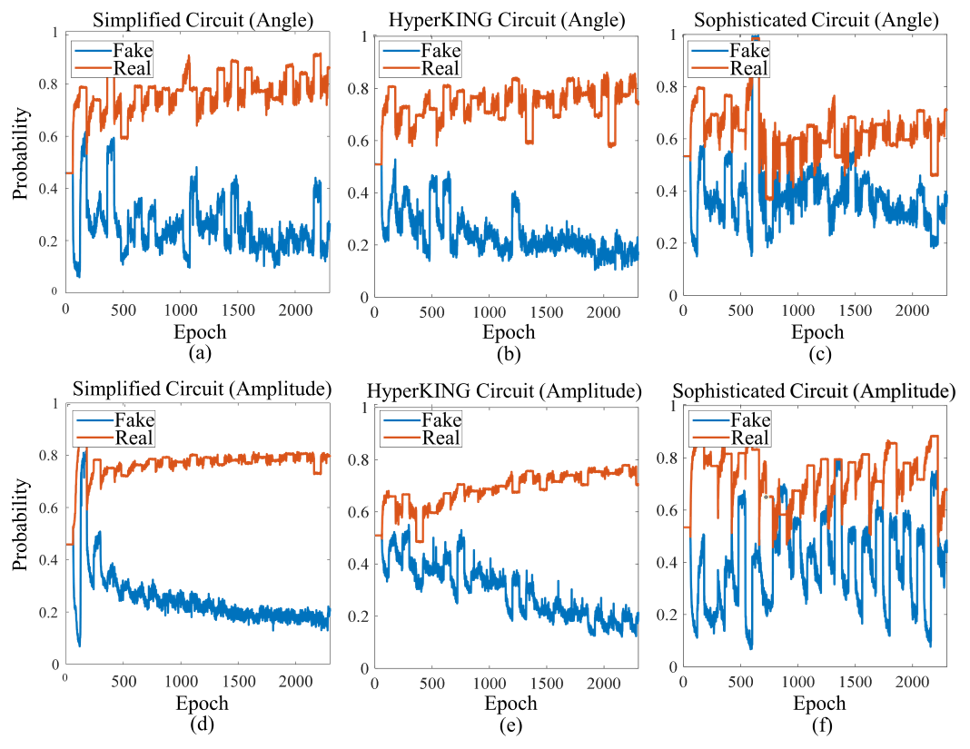}
    \vspace{-0.2cm}
    \caption{The probability curves (defined in Figure \ref{fig:QAL curve}) for the proposed HyperKING framework and its variations, including the more simplified or more sophisticated quantum EM module (the core in our hybrid discriminator).
    We also investigate both the amplitude and angle encoding schemes for better understanding of the {QAL}.}\label{fig:compare_amp}
\end{figure}

\vspace{-0.1cm}
\subsection{Comparability Analysis}\label{subsec:Comparability Analysis}
In this subsection, we show that the proposed hybrid discriminator has comparable ability for adversarially upgrading the hybrid generator.
We also show that this is not a coincidence.
If the core quantum EM module deployed on the hybrid discriminator becomes more sophisticated or more simplified, the adversarial learning may not be smoothly done.
As mentioned in Section \ref{sec:Method}, the function of the hybrid generator $\text{G}(\cdot)$ is to generate a plausible image to cheat the hybrid discriminator, so we naturally adopt the modified entropy loss function \cite{ADMM-Adam}
\begin{align}\label{eq:lossG}
    {\mathcal{L}_{\text{G}}}=\|\mathcal{I}_{\text{real}}-\mathcal{I}_{\text{fake}}^\textrm{(G)}\|_{\text{S}}
+\lambda~\!\log{(1-\text{D}(\mathcal{I}_{\text{fake}}^\textrm{(G)})+\delta)},
\end{align}
where $\| \cdot \|_{\text{S}}$ is the smoothed $\ell$-1 norm \cite{girshick2015fast} to mitigate the outlier effect, $\mathcal{I}_\text{{real}}$ is the real image, $\mathcal{I}_{\text{fake}}^\textrm{(G)}$ is the fake image generated by $\text{G}(\cdot)$, $\lambda:=0.01$, $\delta:=10^{-8}$, and $\text{D}(\cdot)$ is the hybrid discriminator, respectively.
Similarly, we adopt
\begin{align}\label{eq:lossD}
    {\mathcal{L}_{\text{D}}}=-[\log{(\text{D}(\mathcal{I}_{\text{real}})+\delta)}+\log{(1-\text{D}(\mathcal{I}_{\text{fake}}^\textrm{(G)})+\delta)}]
\end{align}
as the entropy loss for training the quantum discriminator.

Accordingly, a successful QAL is expected to have the following attributes.
The hybrid discriminator needs to force the hybrid generator to become more and more stronger.
However, when the discriminator is too strong, the generator may give up to become stronger to be able to cheat the discriminator.
On the contrary, the hybrid generator may not evolve to a stronger state, if the hybrid discriminator is too weak or too easy to be cheated.
Now, we analyze the adversarial curves of HyperKING as displayed in Figure \ref{fig:QAL curve}, which shows a successful QAL.
First, in the zoom-in local region (cf. the dashed red rectangle in Figure \ref{fig:QAL curve}), one can observe the desired separability of the fake and real image curves during the training period of the hybrid discriminator (i.e., the yellow blocks in Figure \ref{fig:QAL curve}), meaning that the discriminator can successfully distinguish whether the input images are real or fake.
Second, during the training period of the hybrid generator $\textrm{G}(\cdot)$ (i.e., the gray blocks in Figure \ref{fig:QAL curve}), the fake image curve increases as expected, since the generator is being trained to become more capable of generating more plausible images to cheat $\textrm{D}(\cdot)$ (cf. \eqref{eq:lossD}).
The real image curve remains nearly constant because $\textrm{D}(\cdot)$ remains unchanged during the training period of $\textrm{G}(\cdot)$.
Beside the above local attributes, from the global range displayed in the upper part of Figure \ref{fig:QAL curve}, one can clearly observe the global separability between the two probability curves, indicating that the QAL successfully helps the discriminator to gradually evolve to stronger states.
In summary, a successful QAL is expected to have the above local and global attributes.
We remark that due to the inevitable randomness like data shuffle, the QAL procedure may not always exhibit these attributes, but under the aforementioned two sets of computer facilities (cf. Section \ref{subsec:Experimental Settings}), about $70\%$ successful rate for the adversarial learning has been observed.
As a side remark for future users of HyperKING (probably with different applications), both the learning rate and the initial seed could also affect the rate of success \cite{picard2023torchmanualseed3407}.

To understand that the successful QAL achieved in Figure \ref{fig:QAL curve} is not easy, we show that such a balance may be broken if $\textrm{G}(\cdot)$ or $\textrm{D}(\cdot)$ gets too strong or too weak.
To this end, the hybrid discriminator is revised to have more sophisticated or more simplified quantum structures, while the classical parts remain unchanged.
Specifically, we enhance the 4-qubit EM module $\textrm{EM}(bcd~\!\!\!\mid~\!\!\!a)$ in the hybrid discriminator (cf. Figure \ref{fig:Dnet}), to the stronger 8-qubit version $\overline{\textrm{EM}}(\bv_{-k}^{(8)}~\!\!\!\mid~\!\!\!k)$ ($k:=0,\dots,7$) and to the weakened 2-qubit version $\underline{\textrm{EM}}(\bv_{-k}^{(2)}~\!\!\!\mid~\!\!\!k)$ ($k:=0,1$), where the notation $\bv_{-k}^{(8)}$ is sequentially composed of all the qubits ``$7,6,5,\dots,1,0$" but ``$k$" (and $\bv_{-k}^{(2)}$ is similarly defined).
For example, we have $\overline{\textrm{EM}}(\bv_{-4}^{(8)}~\!\!\!\mid~\!\!\!4)=\overline{\textrm{EM}}(7653210~\!\!\!\mid~\!\!\!4)$, denoting the quantum module that uses qubit ``$4$" to sequentially control the other seven qubits ``$7653210$" through the CRX gates, as graphically illustrated in Figure \ref{fig:8qubit_circut_enta}.
Accordingly, the more sophisticated quantum structure is designed by replacing the four EM modules of $\textrm{D}(\cdot)$ with the eight modules ``$\overline{\textrm{EM}}(\bv_{-7}^{(8)}~\!\!\!\mid~\!\!\!7)$, $\overline{\textrm{EM}}(\bv_{-6}^{(8)}~\!\!\!\mid~\!\!\!6)$,..., $\overline{\textrm{EM}}(\bv_{-0}^{(8)}~\!\!\!\mid~\!\!\!0)$", and the more simplified structure is to replace the EM modules with the two modules ``$\underline{\textrm{EM}}(\bv_{-1}^{(2)}~\!\!\!\mid~\!\!\!1)$, $\underline{\textrm{EM}}(\bv_{-0}^{(2)}~\!\!\!\mid~\!\!\!0)$".
Thus, there are in total three modules (i.e., $\overline{\textrm{EM}}$, ${\textrm{EM}}$, and $\underline{\textrm{EM}}$) and two quantum encoding schemes (i.e., angle encoding and amplitude encoding), leading to a total of six potential combinations.
All the six combinations have been tested with results collectively summarized in Figure \ref{fig:compare_amp}.

One can see that the original HyperKING architecture [i.e., Figure \ref{fig:compare_amp}(e)], which adopts the amplitude encoder in $\textrm{D}(\cdot)$ (cf. Figure \ref{fig:Dnet}), best holds the desired attributes of a successful QAL, among the six cases.
Comparing to the amplitude encoding scheme [cf. Figures \ref{fig:compare_amp}(d)-\ref{fig:compare_amp}(f)], the angle encoding has relatively weak global separability between the real and fake image curves [cf. Figure \ref{fig:compare_amp}(a)-\ref{fig:compare_amp}(c)].
The reason is that amplitude encoding encodes exponentially more quantum feature information when comparing to the angle encoding \cite{QIPbook}, meaning that amplitude encoding scheme corresponds to a much larger receptive field.
Also, one can see that when the architecture of the quantum module deployed on the discriminator is too strong/sophisticated [cf. Figures \ref{fig:compare_amp}(c) and \ref{fig:compare_amp}(f)], the generator would give up to cheat it, leading to the completely failed QAL.
Though this situation has been improved for the case of simplified discriminator architecture [cf. Figures \ref{fig:compare_amp}(a) and \ref{fig:compare_amp}(d)], the desired curve attributes are not as clear as that of the original HyperKING [i.e., Figure \ref{fig:compare_amp}(e)].
However, one can still see that for the simplified case, the amplitude encoding also exhibits better adversarial learning curves than the angle encoding.

To further emphasize the significance of balanced comparability between the hybrid generator and discriminator, we quantitatively evaluate the performance within the same epoch range under these QALs, as presented in Figure \ref{fig:compare_amp}. 
The details of the landscapes and quantitative metrics used in this experiment can be found in Section \ref{subsec:Quantitative Evaluation}.
Initially, in the cases of the sophisticated discriminators [i.e., Figure \ref{fig:compare_amp}(c) and Figure \ref{fig:compare_amp}(f)], they merely achieve approximately 31dB of PSNR averaged across the landscapes due to the failed QALs.
Conversely, the cases of the simplified discriminators [i.e., Figure \ref{fig:compare_amp}(a) and Figure \ref{fig:compare_amp}(d)] improve performance into a PSNR of around 34dB because their comaprabllity are relatively balanced compared to the sophisticated cases according to the adversarial curves.
This observation is also well aligned with the above analysis for comparability between the hybrid generator and discriminator.
Nevertheless, the simplified architectures may not be effective enough to push our hybrid generator to further upgrade.
In the cases of the proposed HyperKING [i.e., Figure \ref{fig:compare_amp}(b) and Figure \ref{fig:compare_amp}(e)], they further upgrade the PSNR by approximately 1dB compared with those simplified cases.
On the other hand, the discriminator with amplitude encoding can more effectively force the generator to become stronger than the discriminator with angle encoding; for example, the former achieves an SAM index of 2.5 degree, whereas the latter only reaches 3.2.
This is also consistent with our analysis of the adversarial curves.
It is interesting to note that although the adversarial curves presented in Figure \ref{fig:compare_amp}(d) seem to be smoother than our original HyperKING [i.e.,  Figure \ref{fig:compare_amp}(e)], the original HyperKING still accomplishes the best quantitative performance.
For this observation, we remark that the smooth training curves do not necessarily yield the best performance due to the intractable non-convexity of deep learning \cite{batchanalysis}.
Specifically, an over-smooth training may rapidly converge to an unfavorable local minimum, whereas the slight instability does help the network to escape out of some local optimal regions, hence leading to better results.
Also, the simplified discriminator hardly pushes the generator to the optimal equilibrium, as we mentioned.

All in all, the judiciously designed HyperKING has proposed a hybrid discriminator that is comparable to the hybrid generator, and this comparability could be broken when the hybrid discriminator/generator are not carefully balanced.
This emphasizes the exquisite structures behind HyperKING.
Based on the abovementioned analysis, we recommend that future researchers design the generator and discriminator with an equal number of qubits for balanced comparability.
Under the balanced QAL framework, one may consider adopting the amplitude encoding in the discriminator to encode exponentially more quantum feature information, thereby pushing the generator to achieve better results.

\begin{table}[t]
\normalsize
\centering
    \caption{Inference time $T$ (seconds) of the proposed HyperKING and other restoration baselines.}\label{table:peertable_realdata}
    \setlength{\tabcolsep}{0.05mm}{ 
\begin{tabular}{cc|ccccc}
    \hline
    \rule[-1ex]{0pt}{3ex}
    & Methods & ~Time $T$~
    \\
    \hline
    \rule{0pt}{2.2ex}
    &LRTDTV{\cite{8233403}}&{11.920}
    \\
    \rule{0pt}{2.2ex}
    & HDIP{\cite{9022040}}&{16.412}
    \\
    \rule{0pt}{2ex}
    &HTNN-FFT{\cite{9730793}}&{45.307}
    \\
    \rule{0pt}{2ex}		
    &R-DLRHyIn{\cite{10032531}}&{43.475}
    \\
     &HyperQUEEN{\cite{HyperQUEEN}}&{\bf 0.037}
    \\
     &HyperKING&{0.057}
    \\
    \hline
\end{tabular}}
\end{table}

\begin{figure}[t]
    \centering
   \includegraphics[width=0.6\linewidth]{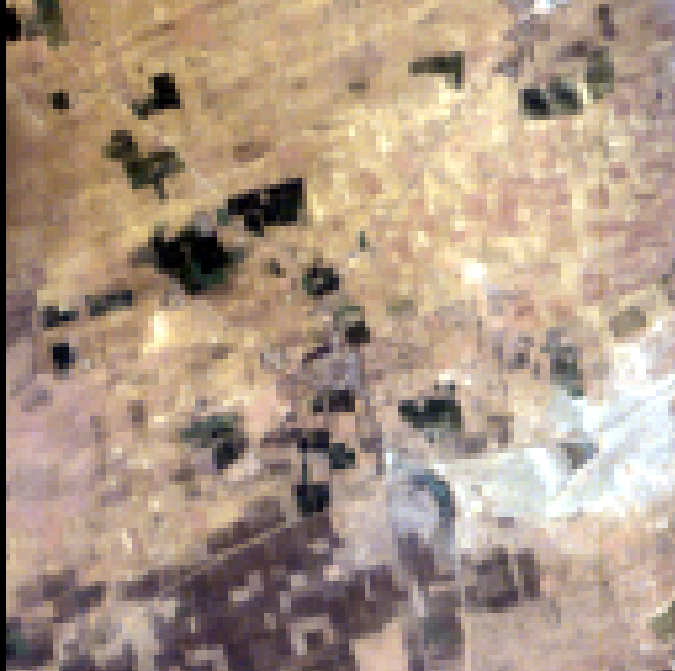}
    \caption{False color composition [bands $(32,23,12)$ as RGB] of the well-known real hyperspectral data over Bhilwara of India.}\label{fig:Bhi}
\end{figure}

\vspace{-0.2cm}

\subsection{Qualitative Evaluation}\label{subsec:Qualitative Evaluation}
\begin{figure*}[t]
    \centering
   \includegraphics[width=1\linewidth]{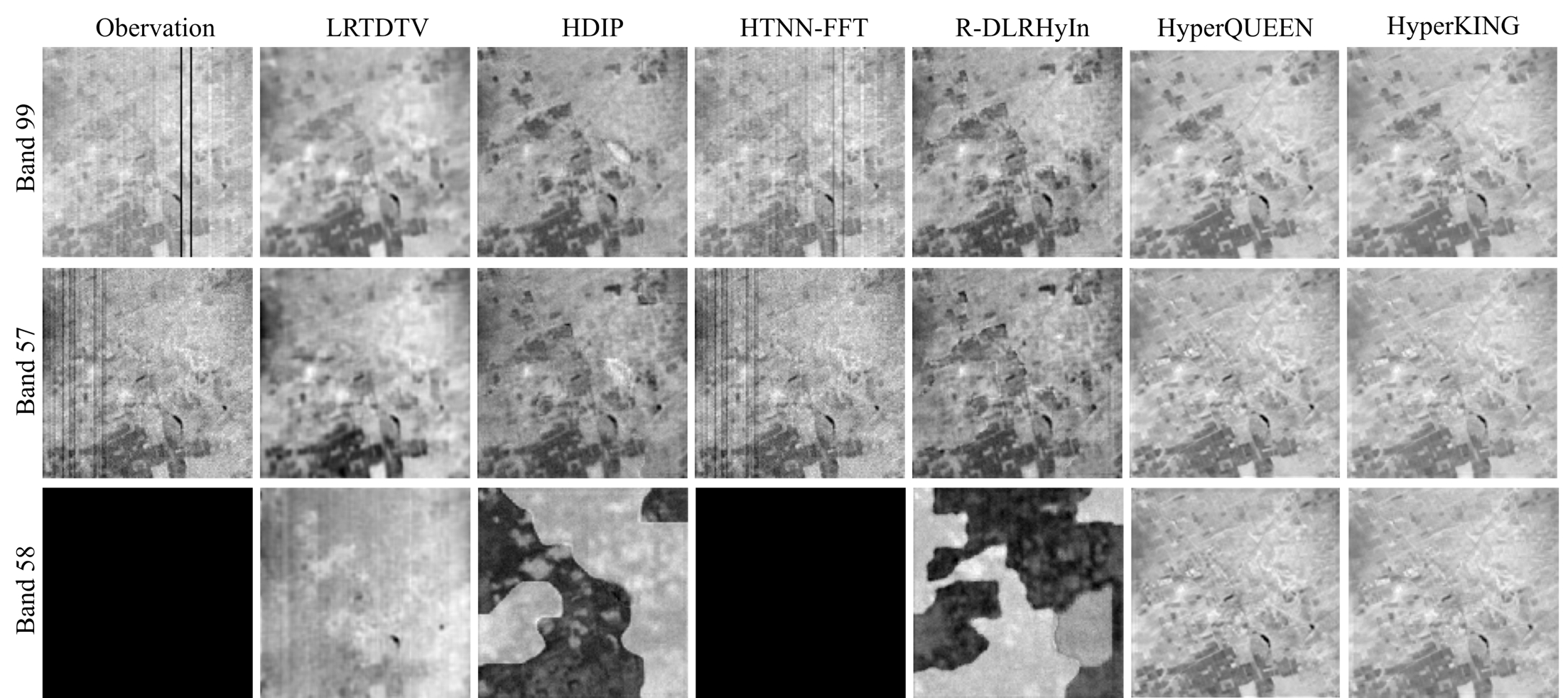}
     \vspace{-0.2cm}
    \caption{Visual comparisons of restored NASA’s EO-1 Hyperion satellite images captured over Bhilwara of India.
    The representative spectral bands regarding sparse corruption (i.e., band 99), dense corruption (i.e., band 57), and serious corruption (i.e., band 58) are leveraged for this qualitative evaluation.}\label{fig:hyperion}
   \vspace{-0.3cm}
\end{figure*}

To investigate the qualitative performance of our quantum-based HyperKING when processing SRS images, we test the renowned real-world hyperspectral image captured by NASA's EO-1 satellite over Bhilwara of India \cite{Bhilwara}.
The sensor captures the wavelengths ranging from 0.4 to 2.5 $\mu$m with a spectral resolution of 10 nm, forming hyperspectral images with 224 spectral bands.
The region of interest (ROI) containing $128\times 128$ pixels \cite{Bhilwaradata} is illustrated in Figure \ref{fig:Bhi}.
We remove those water-vapour bands (i.e., 1-7, 61-77, 122-128, 166-178, 217-242 \cite{puletti2016evaluating}) and use 172 spectral bands in this evaluation.
Besides, we select some representative bands with different types of corruption, including band 100, band 99, band 58, and band 57 (cf. the first column of Figure \ref{fig:hyperion}).

\begin{table}[t]
\footnotesize
\centering
    \caption{The quantitative evaluation of HyperKING and other baselines over city landscape.}\label{table:peertable_city}
    \setlength{\tabcolsep}{0.05mm}{ 
\begin{tabular}{cc|ccccc}
    \hline
    \rule[-1ex]{0pt}{3ex}
    & Methods$~$&$~$PSNR $\!(\uparrow)~$&SAM $\!(\downarrow)~$& RMSE $\!(\downarrow)~$&{SSIM} $\!(\uparrow)~$&{~Time $T$~}
    \\
    \hline
    \rule{0pt}{2.2ex}
    &LRTDTV{\cite{8233403}}~&{28.721}&{5.265}&{0.015}&{0.915}&{21.285}
    \\
    \rule{0pt}{2.2ex}
    &HDIP{\cite{9022040}}&{36.702}&{6.224}&{0.014}&{0.941}&{21.170}
    \\
    \rule{0pt}{2ex}
    &HTNN-FFT{\cite{9730793}}$~$&{30.468}&{5.369}&{0.015}&{0.909}&{36.379}
    \\
    \rule{0pt}{2ex}		
    &R-DLRHyIn{\cite{10032531}}$~$&{37.042}&{6.038}&{0.014}&{0.943}&{48.134}
    \\
    \rule{0pt}{2ex}	
    &HyperQUEEN{\cite{HyperQUEEN}}$~$&{38.918}&{3.557}&{0.008}&{0.979}&{\bf 0.026}
     \\
    \rule{0pt}{2ex}	
    &HyperKING$~$&{\bf 41.971}&{\bf 3.049}&{\bf 0.007}&{\bf 0.984}&{ 0.057}
    \\
    \hline
\end{tabular}}
\end{table}

\begin{table}[t]
\footnotesize
\centering
    \caption{The quantitative evaluation of HyperKING and other baselines over farm landscape.}\label{table:peertable_farm}
    \setlength{\tabcolsep}{0.05mm}{ 
\begin{tabular}{cc|ccccc}
    \hline
    \rule[-1ex]{0pt}{3ex}
    & Methods$~$&$~$PSNR $\!(\uparrow)~$&SAM $\!(\downarrow)~$& RMSE $\!(\downarrow)~$&{SSIM} $\!(\uparrow)~$&{~Time $T$~}
    \\
    \hline
    \rule{0pt}{2.2ex}
    &LRTDTV{\cite{8233403}}~&{29.787}&{\bf 1.977}&{0.017}&{0.929}&{21.039}
    \\
    \rule{0pt}{2.2ex}
    &HDIP{\cite{9022040}}&{28.032}&{5.161}&{0.022}&{0.925}&{16.031}
    \\
    \rule{0pt}{2ex}
    &HTNN-FFT{\cite{9730793}}$~$&{31.568}&{2.071}&{0.015}&{0.940}&{36.550}
    \\
    \rule{0pt}{2ex}		
    &R-DLRHyIn{\cite{10032531}}$~$&{29.204}&{4.500}&{0.020}&{0.935}&{4.514}
    \\
    \rule{0pt}{2ex}	
    &HyperQUEEN{\cite{HyperQUEEN}}$~$&{31.014}&{3.538}&{0.017}&{0.967}&{\bf 0.025}
     \\
    \rule{0pt}{2ex}	
    &HyperKING$~$&{\bf 33.461}&{ 2.802}&{\bf 0.013}&{\bf 0.981}&{0.057}
    \\
    \hline
\end{tabular}}
\end{table}

\begin{table}[t]
\footnotesize
\centering
    \caption{The quantitative evaluation of HyperKING and other baselines over lake landscape.}\label{table:peertable_lake}
    \setlength{\tabcolsep}{0.05mm}{ 
\begin{tabular}{cc|ccccc}
    \hline
    \rule[-1ex]{0pt}{3ex}
    & Methods$~$&$~$PSNR $\!(\uparrow)~$&SAM $\!(\downarrow)~$& RMSE $\!(\downarrow)~$&{SSIM} $\!(\uparrow)~$&{~Time $T$~}
    \\
    \hline
    \rule{0pt}{2.2ex}
    &LRTDTV{\cite{8233403}}~&{30.971}&{3.791}&{0.016}&{0.973}&{21.611}
    \\
    \rule{0pt}{2.2ex}
    &HDIP{\cite{9022040}}&{30.485}&{5.702}&{0.017}&{0.966}&{16.287}
    \\
    \rule{0pt}{2ex}
    &HTNN-FFT{\cite{9730793}}$~$&{28.470}&{\bf 2.880}&{0.014}&{0.956}&{38.312}
    \\
    \rule{0pt}{2ex}		
    &R-DLRHyIn{\cite{10032531}}$~$&{31.939}&{4.226}&{0.013}&{0.974}&{43.904}
    \\
    \rule{0pt}{2ex}	
    &HyperQUEEN{\cite{HyperQUEEN}}$~$&{30.794}&{3.468}&{0.014}&{0.985}&{\bf 0.025}
     \\
    \rule{0pt}{2ex}	
    &HyperKING$~$&{\bf 33.604}&{ 2.899}&{\bf 0.010}&{\bf 0.991}&{ 0.057}
    \\
    \hline
\end{tabular}}
\end{table}
\vspace{-0.02cm}
\begin{table}[t]
\footnotesize
\centering
    \caption{The quantitative evaluation of HyperKING and other baselines over mountain landscape.}\label{table:peertable_mountain}
    \setlength{\tabcolsep}{0.05mm}{ 
\begin{tabular}{cc|ccccc}
    \hline
    \rule[-1ex]{0pt}{3ex}
    & Methods$~$&$~$PSNR $\!(\uparrow)~$&SAM $\!(\downarrow)~$& RMSE $\!(\downarrow)~$&{SSIM} $\!(\uparrow)~$&{~Time $T$~}
    \\
    \hline
    \rule{0pt}{2.2ex}
    &LRTDTV{\cite{8233403}}~&{31.086}&{1.826}&{0.018}&{0.929}&{20.695}
    \\
    \rule{0pt}{2.2ex}
    &HDIP{\cite{9022040}}&{27.133}&{4.692}&{0.031}&{0.854}&{16.227}
    \\
    \rule{0pt}{2ex}
    &HTNN-FFT{\cite{9730793}}$~$&{32.012}&{1.855}&{0.016}&{0.937}&{36.763}
    \\
    \rule{0pt}{2ex}		
    &R-DLRHyIn{\cite{10032531}}$~$&{27.929}&{4.257}&{0.028}&{0.868}&{43.359}
    \\
    \rule{0pt}{2ex}	
    &HyperQUEEN{\cite{HyperQUEEN}}$~$&{32.984}&{1.731}&{0.016}&{0.965}&{\bf0.026}
     \\
    \rule{0pt}{2ex}	
    &HyperKING$~$&{\bf 35.990}&{\bf 1.376}&{\bf 0.010}&{\bf 0.984}&{ 0.06}
    \\
    \hline
\end{tabular}}
\end{table}

The restoration results and inference time are, respectively, presented in Figure \ref{fig:hyperion} and Table \ref{table:peertable_realdata}.
One can observe that band 99 and band 57 have sparse and dense strip missing patterns, respectively.
From Figure \ref{fig:hyperion}, we can see that only HyperQUEEN and the proposed HyperKING have well recovered the corrupted regions without any visual or color distortions.
As for band 58, it is the most serious case corresponding to the completely missing pattern.
In this scenario, HDIP and R-DLRHyIn may have some unexpected spatial patterns because of the absence of a discriminator, while HTNN-FFT fails to recover this band image.
LRTDTV and HyperQUEEN have reasonable recovery results, but LRTDTV does not well align with the ground-truth spatial texture as presented in the ROI (cf. Figure \ref{fig:Bhi}).
The proposed HyperKING exhibits its superiority and well recovers this completely missing band under a fully blind setting.

In summary, the proposed HyperKING achieves the best qualitative performance in the real-world satellite data restoration scenario.
Notably, the inference time of quantum-based baseline (i.e., HyperQUEEN) and our quantum-based generative AI (i.e., HyperKING) is several orders of magnitude faster than other classical baseline methods (cf. Table \ref{table:peertable_realdata}), demonstrating the feasibility of the real-time satellite signal processing via quantum machine intelligence (quantum-based generative AI).
Note that as the experimental settings are exactly the same as in \cite{HyperQUEEN}, the experimental results of HyperQUEEN are replicated from \cite{HyperQUEEN}.
Beyond qualitative evaluation, we will also perform a quantitative evaluation in Section \ref{subsec:Quantitative Evaluation} using all the four representative landscapes (i.e., city, farm, lake, mountain) and Wald’s protocol \cite{ADMM-Adam,wald1997fusion} for a comprehensive comparison.

\begin{figure*}[t]
    \centering
   \includegraphics[width=1\linewidth]{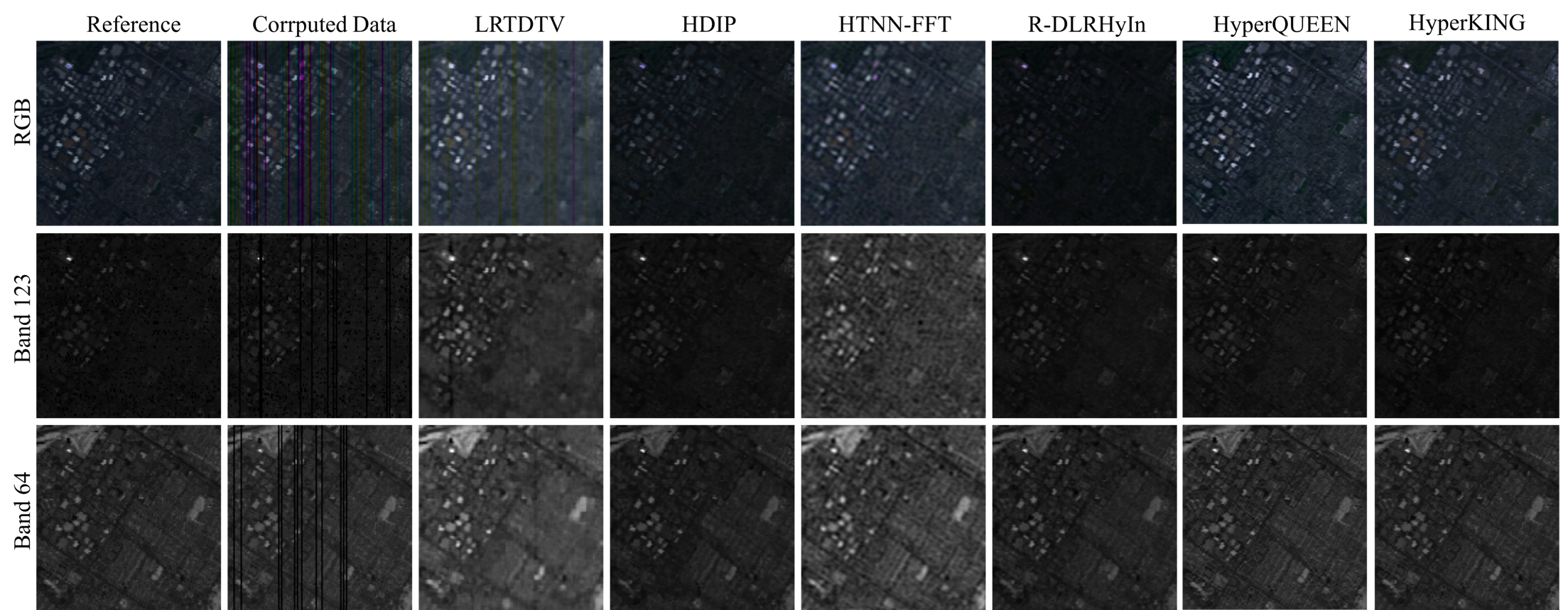}
     \vspace{-0.2cm}
    \caption{Restoration of NASA’s AVIRIS data captured over city landscape. Representative spectral bands, including the true-color composition bands, band 123, and band 64, are displayed.}\label{fig:city}
   \vspace{-0.3cm}
\end{figure*}

\begin{figure*}[t]
    \centering
   \includegraphics[width=1\linewidth]{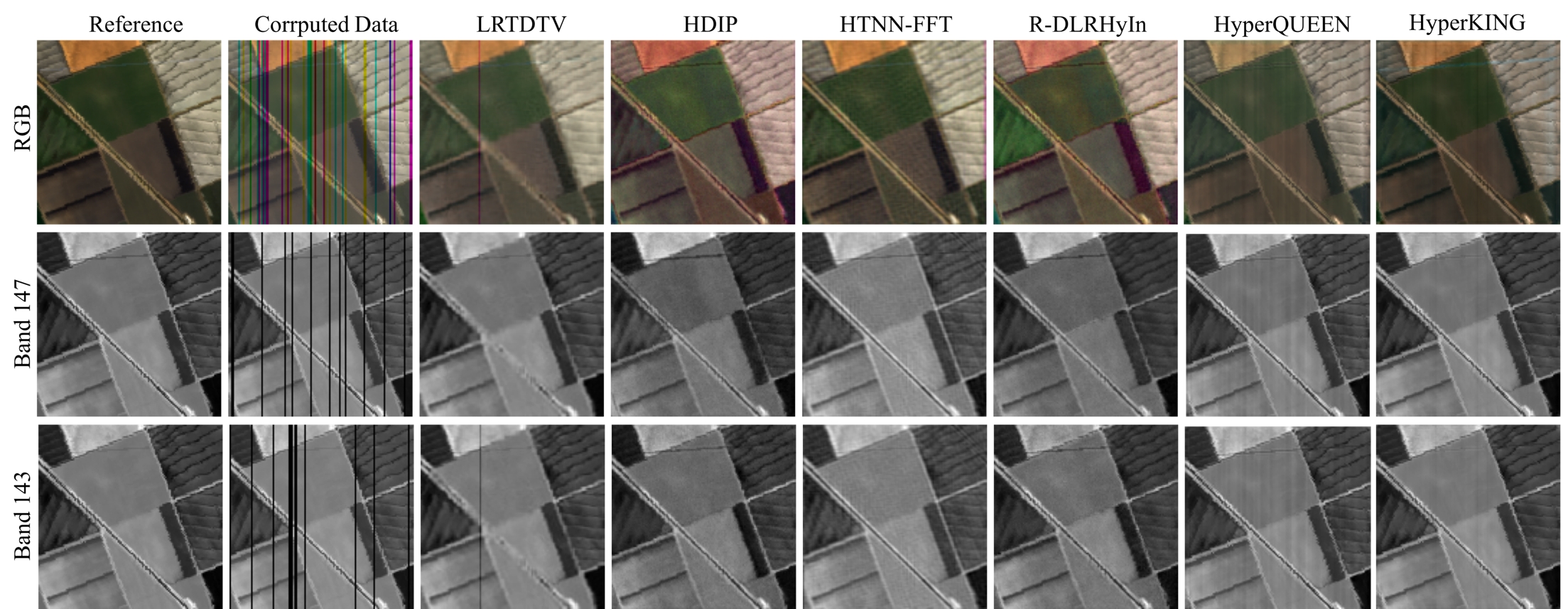}
     \vspace{-0.2cm}
    \caption{Restoration of NASA’s AVIRIS data captured over farm landscape. Representative spectral bands, including the true-color composition bands, band 147, and band 143, are displayed.}\label{fig:farm}
   \vspace{-0.3cm}
\end{figure*}

\begin{figure*}[t]
    \centering
   \includegraphics[width=1\linewidth]{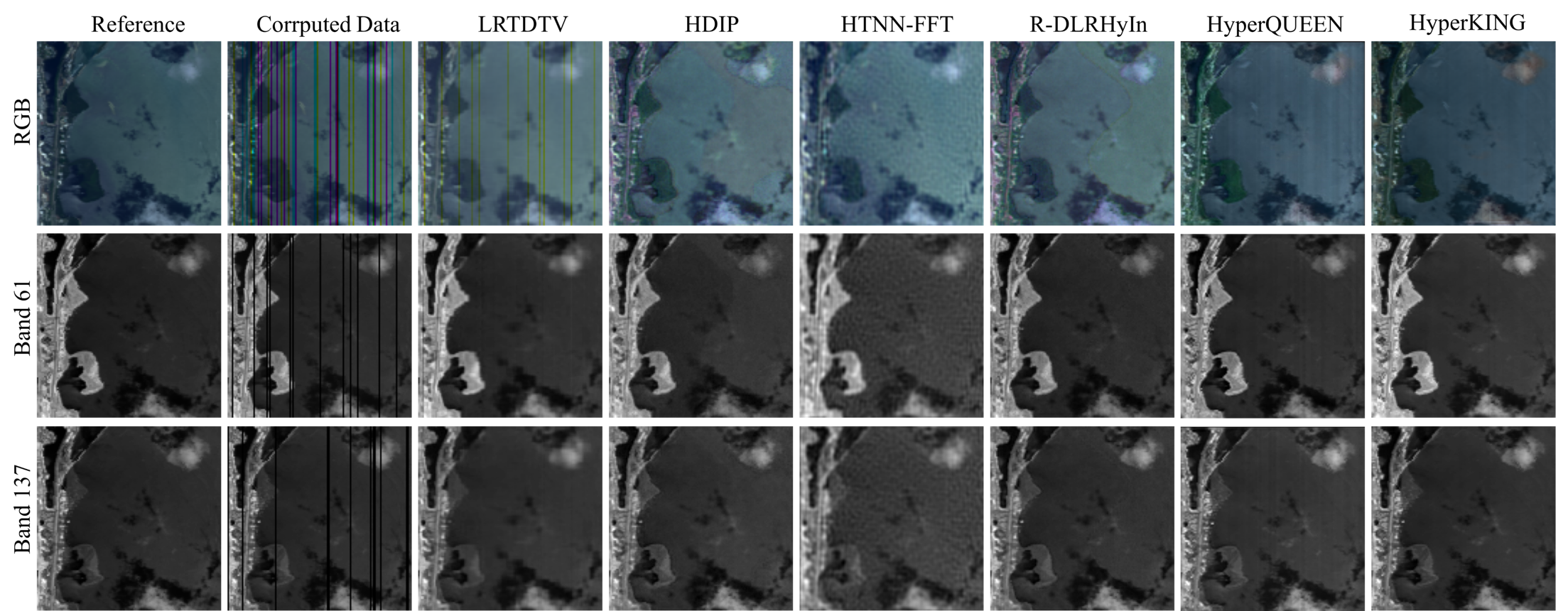}
     \vspace{-0.2cm}
    \caption{Restoration of NASA’s AVIRIS data captured over lake landscape. Representative spectral bands, including the true-color composition bands, band 61, and band 137, are displayed.}\label{fig:lake}
   \vspace{-0.3cm}
\end{figure*}

\begin{figure*}[t]
    \centering
   \includegraphics[width=1\linewidth]{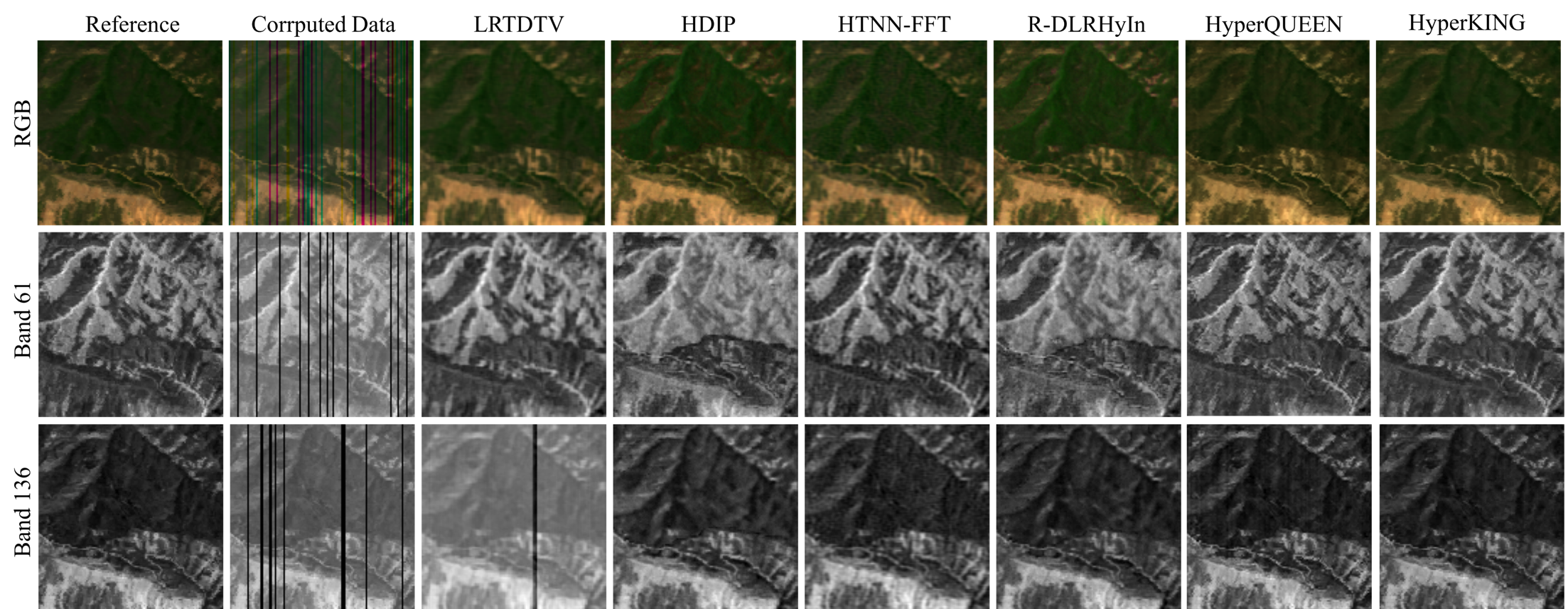}
     \vspace{-0.2cm}
    \caption{Restoration of NASA’s AVIRIS data captured over mountain landscape. Representative spectral bands, including the true-color composition bands, band 61, and band 136, are displayed.}\label{fig:mountain}
   \vspace{-0.3cm}
\end{figure*}
\begin{figure}[t]
    \centering
   \includegraphics[width=1\linewidth]{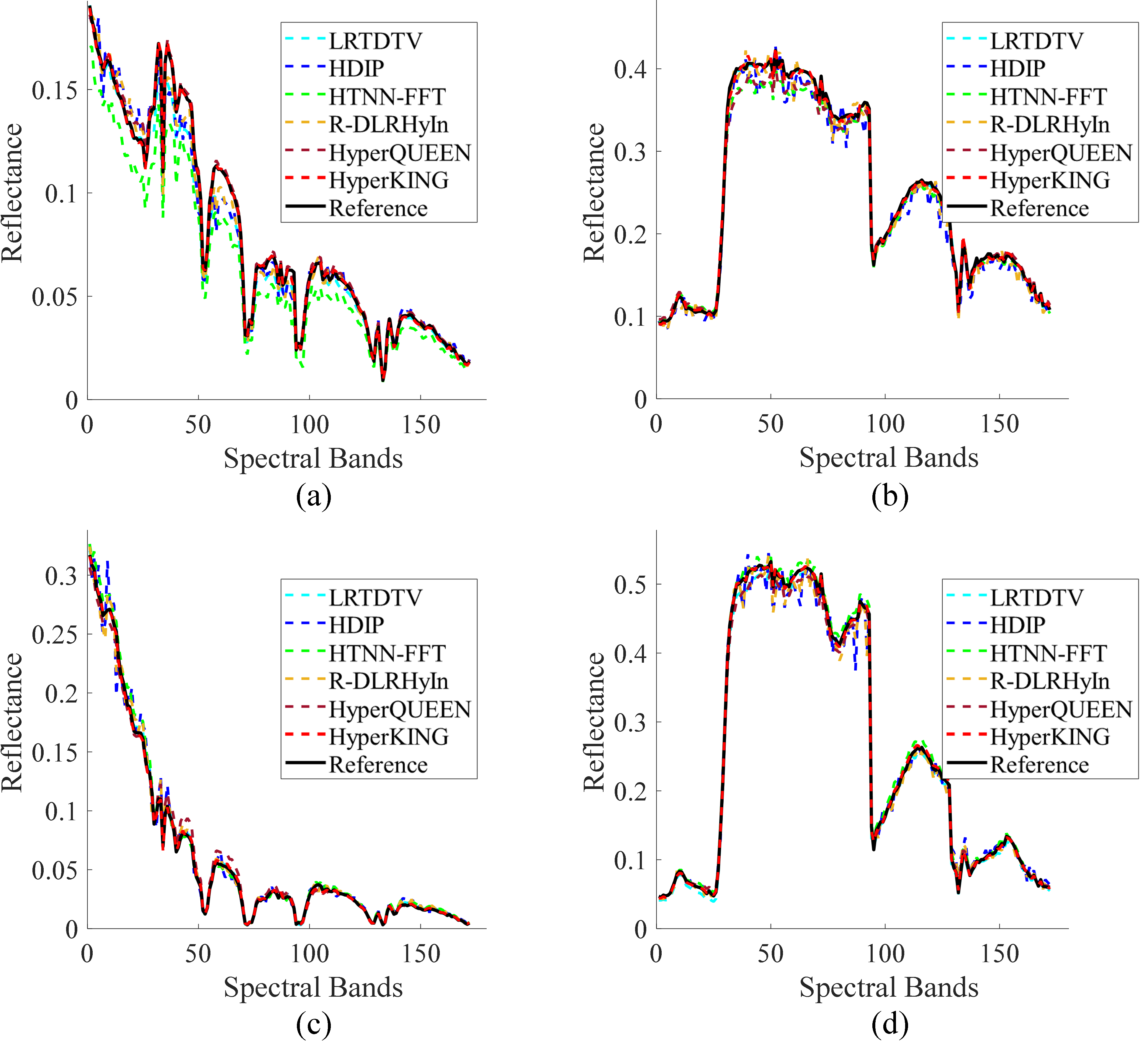}
    \caption{Qualitative comparisons for the spectral profiles over the (a) city, (b) farm, (c) lake, and (d) mountain landscapes.}\label{fig:spectral profile}
\end{figure}

\subsection{Quantitative Evaluation}\label{subsec:Quantitative Evaluation}
Next, we further conduct quantitative evaluations based on well-known quantitative metrics, i.e., peak signal-to-noise
ratio (PSNR) \cite{lin2021all}, spectral angle mapper (SAM) \cite{ADMM-Adam}, root-mean-square error (RMSE) \cite{HyperQUEEN}, and structural similarity (SSIM) \cite{hore2010image}.
Besides, we select the four representative landscapes from NASA’s AVIRIS data (cf. Section \ref{subsec:Experimental Settings}), i.e., city, farm, lake, and mountain, whose ROIs are illustrated in the first column of Figure \ref{fig:city}, Figure \ref{fig:farm}, Figure \ref{fig:lake}, and Figure \ref{fig:mountain}, respectively.
In this evaluation, we employ the dense stripes missing that are widely seen in SRS images to conduct the tensor completion task.
In detail, we randomly remove about 10\% stripes in each spectral band of the corresponding reference (cf. the second columns in Figure \ref{fig:city}, Figure \ref{fig:farm}, Figure \ref{fig:lake}, and Figure \ref{fig:mountain}).
Notably, although the corrupted stripes are padded with zeros, their corrupted patterns differ across each band, resulting in color stripes instead of black stripes in our tensor completion task.
Note that in real-world sensor array systems, color stripes are more frequently seen as the position where the stripe occurs (i.e., the position where the sensor is broken) differs across each band.
Then, the quantitative results of city, farm, lake, and mountain are organized, respectively, in Table \ref{table:peertable_city}, Table \ref{table:peertable_farm}, Table \ref{table:peertable_lake}, and Table \ref{table:peertable_mountain}.

First of all, one can observe that the proposed HyperKING outperforms other baselines over city and mountain ROIs in terms of all the PSNR/SAM/RMSE/SSIM.
Remarkably, the PSNR values over the city and mountain landscapes of HyperKING (cf. Table \ref{table:peertable_city}, and Table \ref{table:peertable_mountain}) are higher than the strongest quantum-based benchmark (i.e., HyperQUEEN) by around 3dB.
This is a direct evidence of the superiority of the QAL.
Furthermore, the SAM and SSIM of the proposed HyperKING are also improved, again demonstrating the feasibility of QAL.
Moreover, the proposed HyperKING also achieves the strongest quantitative performances in terms of PSNR/RMSE/SSIM over farm and lake landscapes (cf. Table \ref{table:peertable_farm} and Table \ref{table:peertable_lake}).
LRTDTV achieves the best SAM in farm landscapes, as it leverages the tensor Tucker decomposition to preserve the inherent global spectral correlation.
On the other hand, HTNN-FFT conducts a high-order convex tensor low-rank model, and this low-rank modeling may lead to such a superior spectral shape preservation, as indicated by the lowest SAM value over lake landscape.
All in all, the proposed HyperKING still exhibits superior performance over all the four representative remote sensing landscapes.
Furthermore, compared with the pioneering quantum-based network, i.e., HyperQUEEN, the performance of HyperKING is significantly enhanced thanks to the QAL.
Also, the inference time for the quantum-based methods shows that real-time satellite data restoration is feasible through quantum-based machine intelligence (e.g., HyperQUEEN) or quantum generative AI (e.g., our HyperKING).

To have better understanding, the corresponding visual results based on these four landscapes are also analyzed.
In Figure \ref{fig:city}, one can observe that the results of the HDIP-based approach (e.g., R-DLRHyIn) have very slight chroma distortion.
On the other hand, LRTDTV restores the corruption with a slight blurring effect, which is similar to the real-world data test on the Bhilwara region (cf. Section \ref{subsec:Qualitative Evaluation}).
Only quantum-based methods (i.e., HyperQUEEN and HyperKING) yield promising visual quality in the city ROI, which echoes the above quantitative analysis (cf. Table \ref{table:peertable_city}).
However, in the farm landscape (cf. Figure \ref{fig:farm}), the results of HyperQUEEN have slight stripe texture, and this has been significantly improved by HyperKING, as the proposed QAL can surge the QUEEN-based hybrid generator to be upgraded to a new level, echoing the result presented in Section \ref{subsec:Comparability Analysis}.
The hybrid generator and hybrid discriminator compete against each other hence achieving even superior performance comparing to the case of independently training a single network.
On the other hand, we remark that HTNN-FFT also achieves promising qualitative performance.
In particular, the upper-right region in the RGB composition of HTNN-FFT is relatively close to the reference, demonstrating its effectiveness in spectral preserving.
However, the unfavorable grid-like patterns in the middle region of HTNN-FFT reveal its limitation in restoring the spatial details.
These observations are well aligned with the quantitative evaluations (cf. Table \ref{table:peertable_farm}); for example, HTNN-FFT achieves promising SAM values, exhibiting its effectiveness in preserving spectral relations.
In lake landscape (cf. Figure \ref{fig:lake}), the slight stripe observed in the result of HyperQUEEN is also eliminated by HyperKING.
As for the mountain landscapes (cf. Figure \ref{fig:mountain}), we can observe that the result of R-DLRHyIn has chroma distortion in the RGB composition while the overall performance is still promising.

On the other hand, preserving the spectral profiles is also critical to facilitate SRS applications.
To this end, the qualitative comparison of the spectral profiles is visualized in Figure \ref{fig:spectral profile}.
We first analyze the spectral profile over the city landscape in Figure \ref{fig:spectral profile}(a), which substantiates that the restored spectral profile of HTNN-FFT has serious deviations from the reference.
Moreover, the spectral profile of other benchmarks (e.g., R-DLRHyIn and HDIP) appears to be noisy, whereas the true spectral profile is expected to be smooth.
This adverse phenomenon is more significant over the farm landscape as shown in Figure \ref{fig:spectral profile}(b).
Notably, only the quantum-based methods (i.e., HyperQUEEN and HyperKING) successfully achieve promising spectral profiles, outperforming the other classical approaches. 
As for the spectral profiles of the lake and mountain landscapes, all approaches seem to have relatively favorable spectral profiles, while HyperKING further suppresses the noisy effect observed in the classical baselines hence accomplishing the most accurate restorations.
All in all, thanks to the effective QAL, the proposed HyperKING has the best qualitative and quantitative performances.

\subsection{Hyperspectral Unmixing Evaluation}\label{subsec:Evidencing Feasibility}
As the key value of hyperspectral image lies within its material identifiability, the restored images should be able to yield correct hyperspectral signatures (a.k.a. endmembers) for the subsequent classification task.
Thus, a critical blind source separation (BSS) technique \cite{BSS1,BSS2} in the SRS field, particularly named hyperspectral unmixing (HU) \cite{HU1}, extracts the endmembers from the target hyperspectral image.
Specifically, a pixel in a remotely sensed hyperspectral image is usually a mixture of several pure endmembers (i.e., material sources).
So, we will need BSS technique to unmix the hyperspectral mixture for recovering the pure endmembers, and this is the so-called HU technique in the SRS area.
We also remark that the HU technique, serving as the fundamental SRS task, empowers numerous SRS applications, such as agricultural monitoring \cite{yu2022critical}, urban planning \cite{8519085}, and anomaly detection \cite{TGFAAD}.
Accordingly, the applicability of the restoration approaches in different SRS applications can be efficiently evaluated by examining whether the restorations from HyperKING can facilitate precise HU, rather than applying them on individual SRS tasks separately.
More specifically, if a restored image can not yield satisfactory HU results, the restoration method is considered less practical, and vice versa.
Therefore, we will use HU result to evaluate the restoration capability of HyperKING, and we will adopt the widely used metric $\phi_{en}$ (i.e., root-mean-square (RMS) spectral angle error) to evaluate the quality of the unmixed endmembers.
A lower value of $\phi_{en}$ corresponds to a better quality of endmembers, and the definition of $\phi_{en}$ can be found in \cite[Equation (32)]{HyperCSI}.

To demonstrate the effectiveness of the proposed QAL, we employ the fast 
 and theoretically reliable HU algorithm, called hyperplane-based Craig simplex identification (HyperCSI) \cite{HyperCSI}.
Specifically, HyperCSI is applied to separate the hyperspectral pixel mixtures for both the reference image and restored image (restored by a particular method like HyperKING), thereby obtaining the original endmembers and the restored endmembers.
Since the endmembers recovered by HyperCSI have been theoretically proven to be the true endmebers under mild conditions, as detailed in \cite[Theorem 2]{HyperCSI}.
Thus, we fairly consider the original endmembers as the reference endmembers, followed by computing the RMS spectral angle error $\phi_{en}$ between the reference and restored endmembers, in order to evaluate the restoration performance.
In the HU algorithm, HyperCSI, the model order $N$, radius compression ratio $r$ and simplex compression ratio $\eta$, are empirically set as $N:=6$, $r:=1E-8$ and $\eta:= 0.9$, respectively.
The HU results are summarized in Table \ref{table:peertable_unmixing}.
As one can see, the proposed HyperKING quantum-based generative AI has achieved the best performance (i.e., the smallest $\phi_{en}$) for all the four investigated landscapes, well echoing the quantitative and qualitative evaluations in Section \ref{subsec:Qualitative Evaluation} and Section \ref{subsec:Quantitative Evaluation}.
With the highly comparable quantum adversarial framework, the hybrid generator well restores the corruptions and maintains rather accurate spectral attributes.
All in all, this BSS experiment again confirms the effectiveness and feasibility of our HyperKING framework.
\begin{table}[t]
\footnotesize
\centering
    \caption{The hyperspectral unmixing (HU) evaluation in terms of $\phi_{en}$ (in degrees) over the four represented landscapes.}\label{table:peertable_unmixing}
    \setlength{\tabcolsep}{0.05mm}{ 
\begin{tabular}{cc|ccccc}
    \hline
    \rule[-1ex]{0pt}{3ex}
    & Methods$~$ & $~~$City $\!~~$& $~~$Farm $\!~~$& $~~$Lake $\!~~$& $~~$Mountain $\!~~$&{~Average~}
    \\
    \hline
    \rule{0pt}{2.2ex}
    &LRTDTV{\cite{8233403}}~&{20.375}&{10.345}&{9.058}&{10.207}&{12.496}
    \\
    \rule{0pt}{2.2ex}
    &HDIP{\cite{9022040}}&{14.311}&{13.537}&{15.086}&{9.123}&{13.014}
    \\
    \rule{0pt}{2ex}
    &HTNN-FFT{\cite{9730793}}$~$&{10.409}&{7.807}&{4.778}&{6.529}&{7.381}
    \\
    \rule{0pt}{2ex}		
    &R-DLRHyIn{\cite{10032531}}$~$&{17.836}&{12.701}&{6.927}&{8.156}&{11.405}
    \\
    \rule{0pt}{2ex}	
    &HyperQUEEN{\cite{HyperQUEEN}}$~$&{9.354}&{12.274}&{5.997}&{7.998}&{8.905}
     \\
    \rule{0pt}{2ex}	
    &HyperKING$~$&{\bf 8.691}&{\bf 5.623}&{\bf 4.056}&{\bf 5.338}&{\bf 5.927}
    \\
    \hline
\end{tabular}}
\end{table}
\subsection{Hyperspectral Mixed Noise Removal Evaluation}\label{subsec:Denoising}
\begin{figure}[t]
    \centering
   \includegraphics[width=0.95\linewidth]{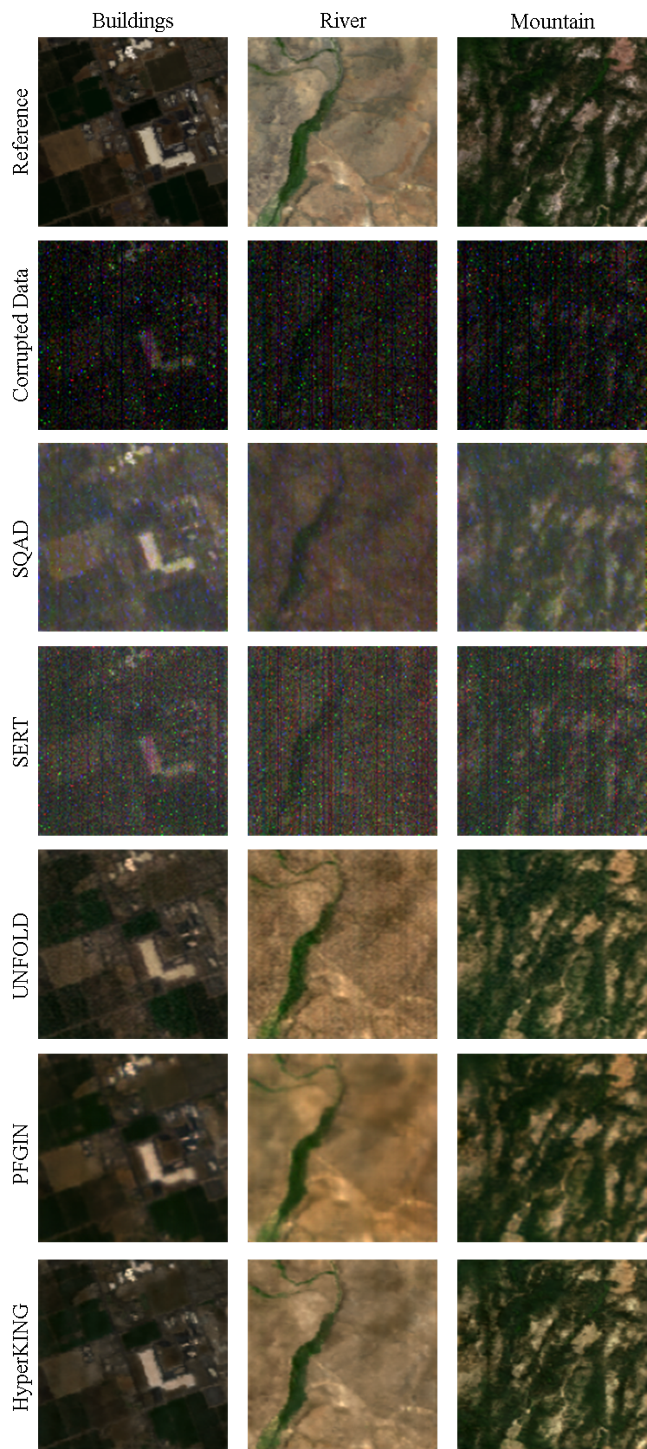}
    \caption{Qualitative comparisons in true-color composition under the severe mixed noise corruption (i.e., dead stripe, Gaussian noise, and impulse noise) over the representative ROIs in the testing dataset.}\label{fig:mix noise}
\end{figure}
In this section, a denoising evaluation is further conducted to substantiate the effectiveness and reliability of the proposed HyperKING.
Given that SRS images are not merely corrupted by the damaged sensor array but also degraded by the unknown complex noise during transportation \cite{SQAD}.
Therefore, removing the noise degradation before those critical SRS applications (e.g., classification) becomes an essential step.
To this end, we adopt a severe mixed noise corruption, including the dead stripes, Gaussian noise, and impulse noise, for this experiment, and the details of mixed noise components are described hereinafter.
First, zero-mean Gaussian noise with a standard deviation being the $10\%$ of maximum reflectance of the HSIs is added for each band.
Subsequently, a total of $1\%$ elements of the HSI are corrupted by impulse noise.
Finally, we randomly add the $10\%$ dead stripe corruption for each band to obtain the mixed noise corrupted HSI.
In summary, each band in the HSI in this experiment is corrupted by three distinct types of noise, leading to a highly challenging degradation, as shown in the second row of Figure \ref{fig:mix noise}.

To evaluate the proposed HyperKING, we adopt various types of supervised baselines for this experiment as the deep learning-based approaches have exhibited their superiority in terms of SRS restoration.
As a result, the selected benchmarks are all supervised.
Initially, we employ the spatial-spectral quasi-attention recurrent
network (abbreviated as SQAD) \cite{SQAD} as the attention-based baseline.
In SQAD, the feature extraction is composed of novel quasi-attention recurrent units for local contextual spatial information and spectral correlation.
With the success of the self-attention mechanism of transformer architectures, the spectral enhanced rectangle transformer for HSI denoising (SERT) \cite{SERT} is also considered as the transformer-based method.
The spatial rectangle self-attention embedded within SERT firstly learns the global dependency of spatial patches, and then the spectral enhancement module extracts the low-rank vector to suppress the noise components. 
Furthermore, an innovative baseline, UNFOLD \cite{UNFLOD}, integrates the strengths of 3D transformer and 3D convolutional neural network (CNN) with a U-shaped architecture. 
Compared to those traditional 2D networks that overlook the 3D textures of HSIs, the 3D transformer and CNN can learn the spectral correlations from the architectural perspective, leading to effective performance.
Finally, we also utilize the very recent classical GAN-based approach, i.e., predictive filtering integrated generative HSI inpainting network (PFGIN) \cite{PFGIN}, as the generative-type baseline for a comprehensive comparison.
All methods are trained with the same training dataset mentioned in Section \ref{subsec:Experimental Settings}.
Besides, we utilize the validation and testing datasets that contain 72 images and 160 images, respectively, for a fair evaluation.
All images in these datasets have a spatial size of 128$\times$128 pixels and are also collected from NASA’s AVIRIS sensor \cite{AVIRISrealdata}.

We conduct the qualitative comparisons as follows.
The representative ROIs selected from the testing dataset, shown in true-color composition, are presented in Figure \ref{fig:mix noise}.
Based on the qualitative comparison, the attention-based approaches, such as SQAD and SERT, fail to effectively restore the HSIs.
In particular, these approaches can only remove noise effects partially due to the challenging noise corruption, leading to unfavorable restoration outcomes.
Next, it can be observed that UNFOLD and PFGIN yield relatively satisfactory restorations, as shown in Figure \ref{fig:mix noise}.
However, the noisy patterns are still visible in the results of UNFOLD,  which leads to spectral deviations (cf. river ROI).
As for the classical GAN-based method, PFGIN, the results seem to be plausible (but over-smoothed) and are lack of spatial details.
Such an inevitable phenomenon reveals the unreliability of classical generative networks.
Compared to these renowned baselines, the proposed quantum-based HyperKING precisely recovers the HSIs from the challenging mixed noise corruption, further demonstrating the effectiveness of QAL.
To further substantiate this fact, we perform the quantitative evaluations as summarized in Table \ref{table:peertable_mixed_noise}.
According to the quantitative evaluation, SERT barely restores the noisy HSI, aligning with the qualitative observations. 
On the other hand, although SQAD accomplishes promising performance in terms of PSNR, its SAM metric is much weaker than UNFOLD and PFGIN.
This spectral deviation is consistent with our qualitative analysis as well.
Conversely, the proposed HyperKING achieves the best quantitative performance among all benchmarks.
In particular, its PSNR metric outperforms the strongest baseline (i.e., PFGIN) by approximately 3dB, substantiating the superiority of QAL. 
All in all, the proposed HyperKING again outperforms the renowned benchmarks under this highly challenging SRS task.
\begin{table}[t]
\footnotesize
\centering
    \caption{The quantitative evaluation of HyperKING and other baselines under the serious mixed noise corruption. The quantitative metrics exhibit the averaged performance over the testing dataset containing 160 HSIs.}\label{table:peertable_mixed_noise}
    \setlength{\tabcolsep}{0.05mm}{ 
\begin{tabular}{cc|ccccc}
    \hline
    \rule[-1ex]{0pt}{3ex}
    & Methods$~$&$~$PSNR $\!(\uparrow)~$&SAM $\!(\downarrow)~$& RMSE $\!(\downarrow)~$&{SSIM} $\!(\uparrow)~$&{~Time $T$~}
    \\
    \hline
    \rule{0pt}{2.2ex}
    &SQAD{\cite{SQAD}}~&{27.387}&{6.536}&{0.025}&{0.761}&{0.095}
    \\
    \rule{0pt}{2.2ex}
    &SERT{\cite{SERT}}&{18.788}&{27.698}&{0.122}&{0.148}&{0.028}
    \\
    \rule{0pt}{2ex}
    &UNFOLD{\cite{UNFLOD}}$~$&{26.854}&{4.431}&{0.019}&{0.828}&{0.031}
    \\
    \rule{0pt}{2ex}		
    &PFGIN{\cite{PFGIN}}$~$&{27.356}&{3.603}&{0.020}&{0.868}&{\bf 0.013}
    \\
    \rule{0pt}{2ex}	
    &HyperKING $~$&{\bf 30.217}&{\bf 2.940}&{\bf 0.016}&{\bf 0.915}&{0.061}
     \\
    \hline
\end{tabular}}
\end{table}

\section{Conclusion and Future Works}\label{sec:Conclusion}

We have proposed the ``Hyperspectral Knot-like IntelligeNt dIscrimiNator and Generator" (HyperKING) quantum framework, which is the first quantum generative AI that achieves advanced quantum image processing tasks in hyperspectral satellite remote sensing.
The hybrid generator in HyperKING is designed as the low-rank quantum deep network (QUEEN) with mathematically provable quantum FE for lightweight quantum AI (cf. Theorem \ref{thm:FE-YXIsing}), which processes the highly compressed quantum features to save the qubit resources in the near-term quantum computer.
The proposed deep compression mechanism is more effective than those naive approaches (e.g., patchwise computing or PCA compression) used in existing benchmark quantum-based GANs.
The QUEEN-based generator also employs the inverse-QC module to map the collapsed quantum image state back to the hyperspectral space.
On the other hand, the hybrid discriminator in HyperKING is designed as the CRX-driven highly-entangled quantum classifier, and has been proven to have comparable ability as the proposed QUEEN-based generator, thereby ensuring that the hybrid discriminator and generator can help each other to get stronger through the quantum adversarial learning.
Achieving the comparable ability has not been easy (cf. Section \ref{subsec:Comparability Analysis}).

The proposed HyperKING leads to a breakthrough in the area of quantum image processing, as it significantly upgrades the spatial capacity (from $28\times 28$ to $128\times128$ resolution) and spectral capacity (from grayscale to 172-channel).
Our HyperKING also achieves superior satellite data restoration results, such as hyperspectral tensor completion and mixed noise removal, with very fast computational time.
Our experiments also show that the hybrid generator/discriminator in HyperKING compete against each other hence achieving significantly superior performance when comparing to the case of independently training a single quantum-based network (e.g., HyperQUEEN).
This is an important reason why we choose GAN as our first step toward quantum generative AI; one would pursue quantum-based diffusion models as the next step.
Remarkably, HyperKING achieves the above objectives under a fully \textit{blind} setting without needing to know the locations of the damaged hyperspectral pixels.
The carefully reported design philosophy behind HyperKING (cf. Section \ref{sec:Method}) also serves as a practical guide for future investigators to study the quantum generative AI.
Another of our ongoing research lines is to explore more advanced applications of HyperKING, such as hyperspectral object counting and hyperspectral classification, for both of which HU-driven identification \cite{EMI2015} is critical.
To this end, we further consider the small-data (or even single-data) quantum adversarial learning as the future work, not only echoing the relatively scarce training samples for SRS classification (comparing to RGB classification) but echoing the limited qubit resources in the near-term quantum computers.
Although our experiment shows that HyperKING trained using very few data (i.e., only 480 hyperspectral images, each with $128\times 128$ pixels) still exhibit high performance for various SRS tasks, its general applicability would be further explored in the future. 
We expect that the proposed HyperKING framework having already proved FE (implying the capability of expressing numerous functionalities for diverse applications) will further achieve other important SRS tasks through just small-data (or single-data) learning, probably requiring customized QAL or upgraded hybrid generator (and its counterpart discriminator with comparable ability; cf. Figure \ref{fig:compare_amp}).

    \bibliography{ref}
    \begin{IEEEbiography}[{\resizebox{0.9in}{!}{\includegraphics[width=1in,height=1.25in,clip,keepaspectratio]{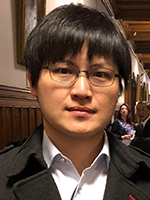}}}]
{\bf Chia-Hsiang Lin}
(S'10-M'18-SM'24)
received the B.S. degree in electrical engineering and the Ph.D. degree in communications engineering from National Tsing Hua University (NTHU), Taiwan, in 2010 and 2016, respectively.
From 2015 to 2016, he was a Visiting Student of Virginia Tech,
Arlington, VA, USA.

He is currently an Associate Professor with the Department of Electrical Engineering, and also with
the Miin Wu School of Computing,
National Cheng Kung University (NCKU), Taiwan.
Before joining NCKU, he held research positions with The Chinese University of Hong Kong, HK (2014 and 2017),
NTHU (2016-2017),
and the University of Lisbon (ULisboa), Lisbon, Portugal (2017-2018).
He was an Assistant Professor with the Center for Space and Remote Sensing Research, National Central University, Taiwan, in 2018, and a Visiting Professor with ULisboa, in 2019.
His research interests include network science,
quantum computing,
convex geometry and optimization, blind signal processing, and imaging science.

Dr. Lin received the Emerging Young Scholar Award (The 2030 Cross-Generation Program) from National Science and Technology Council (NSTC), from 2023 to 2027,
the Future Technology Award from NSTC, in 2022,
the Outstanding Youth Electrical Engineer Award from The Chinese Institute of Electrical Engineering (CIEE), in 2022,
the Best Young Professional Member Award from IEEE Tainan Section, in 2021,
the Prize Paper Award from IEEE Geoscience and Remote Sensing Society (GRS-S), in 2020,
the Top Performance Award from Social Media Prediction Challenge at ACM Multimedia, in 2020,
and The 3rd Place from AIM Real World Super-Resolution Challenge at IEEE International Conference on Computer Vision (ICCV), in 2019.
He received the Ministry of Science and Technology (MOST) Young Scholar Fellowship, together with the EINSTEIN Grant Award, from 2018 to 2023.
In 2016, he was a recipient of the Outstanding Doctoral Dissertation Award from the Chinese Image Processing and Pattern Recognition Society and the Best Doctoral Dissertation Award from the IEEE GRS-S.
\end{IEEEbiography}

\begin{IEEEbiography}[{\resizebox{0.9in}{!}{\includegraphics[width=1in,height=1.25in,clip,keepaspectratio]{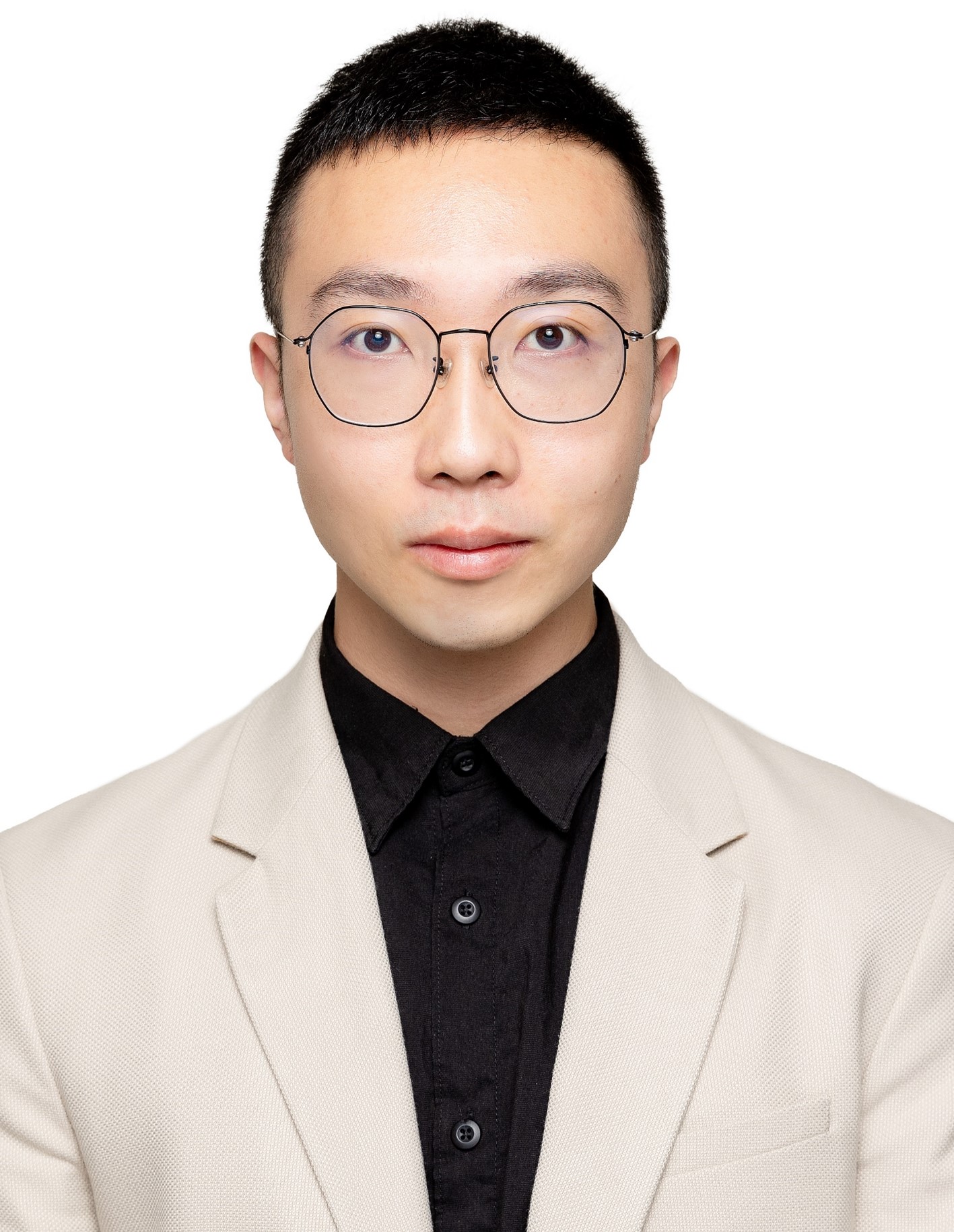}}}]
	{\bf Si-Sheng Young}
(S'23)
is currently a Ph.D. student with the Intelligent
Hyperspectral Computing Laboratory (IHCL), Department of Electrical Engineering, National Cheng Kung University (NCKU), Tainan, Taiwan. 

In 2023, he received the Merit Award from The Grand Challenge ``Computing for the Future", Miin Wu School of Computing, NCKU, as well as the highly competitive ``Pan Wen Yuan Scholarship" from the Industrial Technology Research Institute, Hsinchu, Taiwan.
In 2024, he received a highly competitive ``Scholarship Pilot Program to Cultivate Outstanding Doctoral Students" from the National Science and Technology Council (NSTC), Taipei, Taiwan.
His research interests include convex optimization, deep learning, anomaly detection, data fusion, and imaging inverse problems.
\end{IEEEbiography}

\end{document}